\documentclass[11pt,a4paper]{article}
\usepackage[a4paper, lmargin=0.1666\paperwidth, rmargin=0.1666\paperwidth, tmargin=0.1111\paperheight, bmargin=0.1111\paperheight]{geometry} 
\usepackage[comma]{natbib}
\usepackage{url}
\usepackage{tikz,xcolor}
\usetikzlibrary{trees,arrows,calc}
\usetikzlibrary {decorations.pathmorphing,decorations.pathreplacing,decorations.shapes}
\usepackage{amsmath}
\usepackage{amsfonts}
\usepackage{ulem}
\usepackage{framed,tcolorbox}
\usepackage{hyperref} 
\hypersetup{colorlinks=true,
    allcolors=blue}
\begin{document}
\title{What are Asset Price Bubbles?\\A Survey on Definitions of Financial Bubbles}
\author{Michael Heinrich Baumann$^1$ and Anja Janischewski$^{2,*}$
\\[1eM]
\today\\[3eM]
\tiny$^1$Department of Mathematics, University of Bayreuth, Bayreuth, Germany, D-95447;\\[-.5eM]
\tiny email: michael.baumann@uni-bayreuth.de\\[-.5eM]
\tiny$^2$Department of Economics and Business Administration,\\[-.5eM]
\tiny Chemnitz University of Technology, Chemnitz, Germany, D-09126;\\[-.5eM]
\tiny email: anja.janischewski@wiwi.tu-chemnitz.de\\[-.4eM]
\tiny$^{*}$Corresponding author
}

\date{}

\maketitle

\subsection*{Abstract} 
Financial bubbles and crashes have repeatedly caused economic turmoil notably but not only during the 2008 financial crisis. However, both in the popular press as well as scientific publications, the meaning of bubble is sometimes unspecified. Due to the multitude of bubble definitions, we conduct a systematic review with the following questions: What definitions of asset price bubbles exist in the literature? Which definitions are used in which disciplines and how frequently? We develop a system of definition categories and categorize a total of 122 papers from eleven research areas.

Our results show that although one definition is indeed prevalent in the literature, the overall definition landscape is not uniform. Next to the mostly used definition as deviation from a present value of expected future cash flows, we identify several other definitions, which rely on price properties or other specifications of a fundamental value. This research contributes by shedding light on the possible variations in which bubbles are defined and operationalized. 

\subsection*{Keywords}asset price bubble, fad, financial crisis, local martingale, fundamental analysis

\newpage
\section{INTRODUCTION}\label{sec_int}

When one reads terms like the dot-com bubble, the Tulipmania, the Black Thursday in 1929, or the 2008 financial crisis, one automatically thinks of financial bubbles and crashes. Everyone has a vague idea in her mind of what financial bubbles and crises are, although these definitions tend to be exemplary, possibly stimulated by the popular press. 

Even in financial science, there is no clear definition of a financial bubble. It starts with the question whether a single financial asset exhibits a bubble or whether the whole economy is a so-called bubble economy or---somehow in between---the bubble relates to a sector or industry. Since ``bubbles'' are a multifold problem, a whole bundle of terms has evolved. For example, following \citet[][p.~17]{Kindleberger:78}, ``mania'' emphasizes irrational behavior, ``bubble'' hints at the following crash, and ``crisis'' is determined by
\begin{quote}`speculation, monetary expansion, a rise in the prices of assets followed by a sharp fall, and a rush into money'\end{quote}\citep[][p.~22]{Kindleberger:78}. We observe that the ``rise and fall'' refers to and emphasizes price properties, while ``speculation'' gives a hint to the concept of fundamentals. Using only price path properties, a bubble is defined for example via \begin{quote}`[\dots] an upward price movement over an extended period [\dots] that then implodes'\end{quote}\citep[][p.~25]{Kindleberger:05}.\footnote{Note that, we were not able to find this quote in the very first edition: \citet[][]{Kindleberger:78}. Confer also \citet[][]{Kindleberger:96,Kindleberger:23}.}

Often, economists prefer to use definitions that refer to the so-called fundamental value of a stock or an other financial product, e.g., \citet[][p.~25]{Kindleberger:05} says 
\begin{quote}`Economists use the term bubble to mean any deviation in the price of an asset or a security or a commodity that cannot be explained in terms of the `fundamentals'.'\end{quote}

It is both ambiguous to define fundamental values and difficult to calculate them (using either prices and dividends or balance sheets resp.\ retail values). If one follows the hypothesis of efficient markets, a stock price should always correspond to its fundamental value \citep[cf.][]{Malkiel:89,Malkiel:05,Fama:70}. 
If it deviates, this is called a financial bubble. 

The idea of defining bubbles with the help of fundamental values is quite plausible. A rational trader, if the market is functioning efficiently, will not be willing to pay such a high price that she cannot expect future payoffs to justify that price resp.\ to pay more than the resources in the company that she owns proportionately are worth. However, fundamentals are often non-observable and estimates can be subjective. If a price is above its fundamental value, this does not mean that it will necessarily fall, there is also no reason for it not to \citep[see][]{Barlevy:07}, i.e., a crash can happen. Hence, in stochastic analysis, it may no longer be a price path that is a bubble, but the whole process has a ``bubble property'' inherent in the whole dynamic \citep[see][]{Jarrow:16,Siegel:03,Protter:16}.

While many reviews of bubbles exist in general, including typologies arising from mechanisms that generate bubbles, to our knowledge, there has been no attempt to systematize the different types definitions of bubbles. We clarify the potpourri of definitions by developing a framework with eleven main definition categories and by conducting a systematic review using the search engine Scopus with a final data set of 122 papers. We find that although the majority of papers defines bubbles as deviations of fundamental values there are also authors who only refer to price path properties such as a boom and bust to define or operationalize a bubble. Furthermore, we identify several categories for fundamental value definitions and operationalizations, some of which may be considered equivalent, others not.
Understanding the multitude of bubble definitions is essential when comparing empirical bubble tests or for assessing different regulatory measures for their effectiveness in preventing or mitigating bubbles if they are based on nonequivalent bubble definitions respectively. 

The paper is organized as follows: Section~\ref{sec_lit} provides further motivation and literature context, as well as the research questions. In Section~\ref{sec_cat}, we delineate definitional categories of bubbles as they can be found in the literature and describe selected application areas. Method and results of the systematic review are presented in Section~\ref{sec_rev}. Section~\ref{sec_con} concludes.

\section{Literature and Motivation}\label{sec_lit}
To illustrate diversity of bubble definitions, we present a small selection of quotes below, before we continue to embed our current work in the wider literature on bubbles (Sections~\ref{sec_o_reviews}), discuss theoretically possible types of asset price dynamics (Section~\ref{sec_exdyn}) and clarify the research questions (Section~\ref{sec_questions}). An example why different bubble definitions may not only be interesting for theoretical purposes but leading to real-world questions is given in the last paragraph of Section~\ref{sec_disc}.

\subsection{Motivational quotes}\label{sec_Quotes}
Despite all the ambiguities and the great variety of bubble (and fundamental value) definitions, one finds statements time and again saying that in principle there is indeed a single definition. 
Or, at least, that ``most economists'' prefer a particular (type of) definition \citep[cf.][]{Barlevy:07}.
 
The following quotes might give an overview of how far the ambiguity goes and how unknown it is. 
\citet[][p.\ 46]{Barlevy:07} writes 
\begin{quote}`In particular, most economists would define a bubble as a situation where an asset's price exceeds the ``fundamental'' value of the asset. [\dots]'\end{quote} 
and in the same fashion \citet[][p.\ 88]{Carter:11} say 
\begin{quote}`Most economists define an asset bubble as a period when prices are driven by trader beliefs peripheral to underlying supply and demand factors.'\end{quote} and further \citet[][p.\ 923]{Friedman:09} write \begin{quote}`Modern financial economists define the fundamental value $V$ of an asset as the expected present value, given all available information, of the net cash flow the asset generates. The accepted definition of a bubble is a deviation of market price $P$ from $V$. Crashes are episodes when $B = P - V$ rapidly decreases from a positive value to a zero (or negative) value. 
Beyond these simple definitions, consensus is elusive.'\end{quote} 
However, \citet[][p.\ 45]{Girdzijauskas:09} say 
\begin{quote}`An economic bubble is the commonly used term for an economic cycle that is characterized by a rapid expansion followed by a dramatic crash.'\end{quote} 
while \citet[][p.\ 1229]{Brunnermeier:13} write
\begin{quote}
`The term bubble refers to large, sustained mispricings of financial or real assets. While
definitions of what exactly constitutes a bubble vary, it is clear that not every temporary
mispricing can be called a bubble.'
\end{quote}
Furthermore, Eugene Fama once stated in an interview with \citet{Clement:07}
\begin{quote}`The word ``bubble'' drives me nuts. For example, people say ``the Internet bubble.'' Well, if you go back to that time, most people were saying the Internet was going to revolutionize business, so companies that had a leg up on the Internet were going to become very successful.'\end{quote}
and Rober~J.\ \citet[][p.\ 1487f]{Shiller:14} brought to the point where our research starts: \begin{quote}`There is a troublesome split between efficient markets enthusiasts (who believe that market
prices incorporate accurately all public information and so doubt that bubbles even exist) and
those who believe in behavioral finance (who tend to believe that bubbles and other such
contradictions to efficient markets can be understood only with reference to other social sciences
such as psychology). I suspect that some of the apparent split is illusory, deriving from the problem that
there is not a widely accepted definition of the term ``bubble.'''\end{quote} 
These various quotes on bubble definitions motivates our work to sort and count them.

\subsection{Other bubble reviews}\label{sec_o_reviews} 

There have been several reviews on financial bubbles so far that give an overview over some definitions of bubbles. \citet{Camerer:89} provides a review on the earlier literature on bubbles, sorting the literature according to three drivers of financial bubbles: rational bubbles, fads, and information bubbles. Furthermore, this paper relates the theoretical models to empirical tests for bubbles. Fads describe temporary changes in utility which cause an increase in demand for a certain asset, whereas literature in information bubbles includes (among other topics) models in which agents have heterogeneous information or heterogeneous beliefs about the financial market and economy. A review specifically on rational bubbles is given by \citet{Diba:88}. A review about bubbles in theoretical models with a focus on mathematical finance can be found in \citet{Jarrow:15}. \citet{Jarrow:15} provides a general framework for defining asset price bubbles as deviations from fundamental values combining infinite time horizons and finite time horizons as well as discrete time and continuous time models. The notion of differential beliefs and individual behavior is briefly summarized by  \citet[][p.\ 213]{Jarrow:15}. A brief overview of different possibilities of defining bubbles in the context of non-uniform beliefs among traders is given by \citet{Barlevy:07}. \citet{Kubicova:11} provide a classification of bubbles into four categories: (1) rational investors and identical information, (2) rational investors with asymmetrical information, (3) limits to arbitrage, rational and irrational investors, as well as (4) heterogenous beliefs of investors about fundamental values. \citet{Brzezicka:20} provides a typology of housing price bubbles, in \citet{Kyriazis:20} one finds a systematic review about cryptocurrency bubbles. \citet{Gurkaynak:08} conducts a survey of bubble tests focussing on rational bubbles. A survey on behavioral finance literature including ``irrational bubbles" is provided by \citet{Vissing:03}. Last but not least, \citet{Brunnermeier:13} provide an overview of different mechanisms causing the booms as well as busts in asset prices.

Definitions of bubbles are used, for example, in papers that examine what conditions must be met in an economy for bubbles to occur in general \citep[cf.][]{Samuelson:58,Diamond:65,Tirole:85,Caballero:06,Farhi:12,Rocheteau:11,Allen:93,Conlon:04,Doblas:12,Abreu:03,Blanchard:82}. Most prominently, \citet{Tirole:82} showed that at least one of the following requirements has to be fulfilled for a bubble to occur when traders act rational: there have to be infinitely many traders, traders do not share common beliefs, or there is already some inefficiency incorporated in the economy before the traders start to trade.

While in the already existing surveys on bubbles, different types of bubbles are distinguished, for example, characterizing them via different causal mechanisms, none of the above mentioned reviews provide a systematic overview over the definitions of bubbles. We contribute to the literature in three ways: First, we bridge different bubble definitions as used in mathematics, experimental economics, daily press, finance, etc. Second, we distinguish bubble definitions not according to how bubbles and their formation can be explained, as in the literature above, but according to the definitions themself and discuss the ideas and philosophy behind these definitions as well as which definition types occur in which applications. 
Third, we provide a systematic review to quantitatively verify which definitions are (how often) used.
\vspace{1cm}
\subsection{Exemplary dynamics}\label{sec_exdyn}
\begin{figure}[h]
 \centering
  \includegraphics[width=\textwidth]{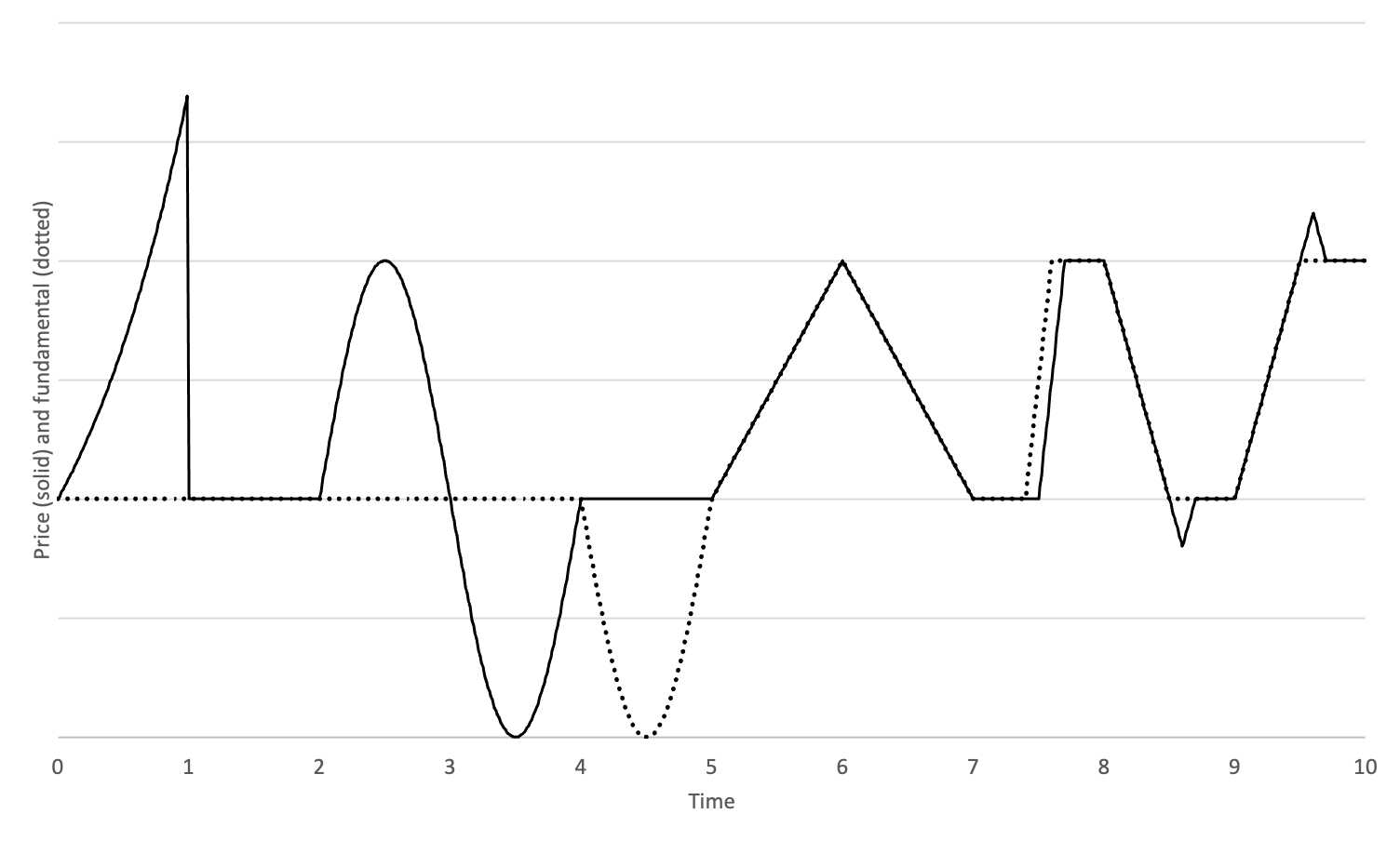}
 \caption{Exemplary paths of an asset's price process and its corresponding fundamental value. Price path with solid lines and fundamentals dotted. (Drawn with MS Excel)}
 \label{fig_Bsp}
\end{figure}

When we look at Figure~\ref{fig_Bsp}, the development between $t=0$ and $t=1$ is what ``everyone'' would call a bubble and the development between $t=1$ and $t=2$ is what would not be a bubble. However, is the development between $t=2$ and $t=3$ a bubble or is the divergence not strong enough or the rise and fall not steep enough? Are downward deviations (between $t=3$ and $t=4$) also bubbles? In economics, this phenomenon is called a negative bubble---but often not considered further \citep[see][]{Barlevy:07}. Between $t=4$ and $t=5$ we see the typical case, which is a bubble if we refer to fundamental values, but no-one if we only look at the price. And this case becomes especially complicated when we consider that fundamental values are unobservable and difficult to estimate. Further, price increases (between $t=5$ and $t=6$) and decreases (between $t=6$ and $t=7$) might not be called bubbles and crashes when referring to fundamentals if these movements are justified by movements of the fundamentals---some would call these price/fundamental value movements fads and corrections. From a practical point of view, another difficulty when referring to fundamentals can be observed between $t=7$ and $t=10$ when prices are delayed or overshoot.

Furthermore, it is important to keep in mind that there are also assets without monetary payoffs such as dividends or similar. For example, companies may not pay out any dividends and keep everything in the company. Then, the fundamental value of these stocks cannot be determined by expected future dividend payments, but needs other estimations of the intrinsic value of the stock shares, for example, the estimation of possible future earnings of the company. Additionally, there is the question of whether money---irrespective of whether it is fiat money or gold(-backed)---is a bubble. Based on these points, it is evident that a uniform definition of bubble is rather difficult. A potpourri of definitions may make more sense, so that a suitable definition can be found for each application.

\subsection{Research questions and restrictions}\label{sec_questions}
We make an attempt in this paper to quantify what ``most'' economists are and whether there are any ``commonly accepted'' definitions at all. To do this, we conduct a systematic review and describe the procedure 
in detail.

In summary, our research questions are the following:
\begin{itemize}
\item What definitions of asset price bubbles exist in the literature?
\item Which definitions are used in which disciplines?
\item Which definitions are used and how frequently? 
\end{itemize}
As a proxy for a representative subset of bubble literature, we choose a subset of literature on Scopus using specified search criteria described in Section~\ref{sec_method}.
Depending on the area of applications, different definition types occur, for example different bubble tests use either only asset prices or both, asset prices and fundamental data. We provide an overview which application area uses which definitions for the areas bubble tests, heterogeneous agent models, experimental economics, housing and real estate prices, history of economics, machine learning, and policies (Section~\ref{sec_appl}). This study is not comprehensive. In detail, we point out that this study is not comprehensive due to the selective but systematic search criteria restrictions, see Section~\ref{sec_method}.

Since our goal is to investigate which bubble definitions exist and are in use, we first have to define the term definition. Since not all papers contain a very clear definition, we apply a certain mix of extracting the definition and the operationalization of a (not explicitly stated) definition. Thus, our process of identifying the definition is the following. i) If there is any phrase like ``We define bubbles as \dots'' the case is accepted. If there is ii) any phrase like ``We use the bubble definition of Citation Year.'' we count the bubble definition of this reference. If iii) the first two points do not apply but a method, methodology, framework for assessing bubbles is used, we take the operationalization of the bubble concept in that method, methodology, or framework as definition. This may in some instances be identical to the underlying definition, in some instances, an operationalization may lead to a paper being categorized in a new bubble definition category. Last, iv) if none of these are true, we try to figure out from the context what the underlying definition is---when this is also not successful, we have to accept that there is v) ``no definition.'' 

Please note that there are various (hotly) debated discussions of some terms and equivalences connected to the issue of financial bubbles. For example, under some neoclassical assumptions bubble definitions---for details of the following terms, consult Section~\ref{sec_cat}---which build upon a) future discounted cash flows, b) resale value, c) equilibrium prices, d) book value, e) utility may be equivalent. However, under other assumptions this may not be true. It is far beyond the scope of our work to discuss these debates, hence, we state and count ``which definitions are used'' irrespective of whether those are equivalent to some others or not. The same is true for equivalences of fundamental properties and price properties, see Sections~\ref{sec_def_fund} and~\ref{sec_def_price}. And last but not least, there is an ongoing discussion of market efficiency---with questions like: is there a chance that a bubble occurs in an efficient market.\footnote{Confer \citet{Fama:13} and \citet{Shiller:14}.} 
We do not get in this discussion, either.

\section{CATEGORIZATION OF BUBBLE PAPERS}\label{sec_cat}
Before describing the systematic review methodology and results in Section~\ref{sec_rev}, we provide an overview of the definition categories we identified in the literature along structural differences (Section~\ref{sec_def_struct}) as well as provide insights on areas of application (Section~\ref{sec_appl}). 

\subsection{Definitions of bubbles}\label{sec_def_struct}

Bubble definitions---as found and applied in the scientific literature---are not homogenous. Here, we describe the categories of bubble definitions along structural differences that we find in the systematic review (Section~\ref{sec_rev}). The main categories are definitions that refer to fundamental values (seven categories) and definitions that refer to properties of the asset price such as a boom and bust (three categories), and a third (small) category of extensional bubble definitions.

\subsubsection{Deviation from fundamental values}\label{sec_def_fund}
A frequently used definition is that 
\begin{tcolorbox}
a bubble is a price that deviates from the fundamental value of the asset. 
\end{tcolorbox}
However, the conceptualization of fundamental values is not homogeneous across the literature. In the following, we describe the various concepts of fundamental values that appear in the bubble literature. As explained above, we do not fully address the potential equivalence of these definitions.

\paragraph{Present value of future expected cash flows}\label{par_cashflow}
Among (many) others, \citet{Summers:86}, \citet{Lee:99}, \citet{Barlevy:07},  and \citet{Jarrow:15} define that 
\begin{tcolorbox}
the fundamental value of financial products is the present  value (i.e.\ with discounting) of all future expected dividends/cash flows,
\end{tcolorbox}
that is,  
``up to eternity.'' The fundamental value (FV) $f_t$ at a time $t$ in a discrete time setting is given by 
\begin{equation}\label{eq:definition_FV}
 f_t = \mathbb{E}\left[\sum_{k = 1}^{\infty} q_{t+k}d_{t+k}\ \middle|\ \mathcal{F}_t\right] \frac{1}{q_t},
\end{equation}
where $q_k$ is a (stochastic) discount factor from time step $k$ back to time step $0$ and $(d_t)_t$ is the stochastic process of cash flows, such as  dividend payments. The expectation is taken w.r.t.\ the set of ``present information" $\mathcal{F}_t$. Under certain assumptions, this formula can be simplified, known as the Gordon Growth model.

The process $f_t$ as defined above is the unique solution to the equation
\begin{equation}\label{eq:price-equation}
p_t = \frac{1}{q_t}\mathbb{E}\left[q_{t+1}\cdot\Big(p_{t+1} + d_{t+1}\Big)\ \middle|\ \mathcal{F}_t\right],
\end{equation}
when solved for $p_t$ such that the transversality condition  
\[\lim_{T \to \infty} \mathbb{E}\frac{1}{q_t}\left[q_{t+T}p_{t+T}\ \middle|\ \mathcal{F}_t\right]= 0\] is fulfilled \citep[see, e.g.,][p.\ 1232]{Brunnermeier:13}. The transversality condition ensures that the present value of the future asset price converges to zero. The definition of the fundamental value depends on the choice of the probability measure in Equation~\eqref{eq:definition_FV}, as further discussed in Section~\ref{sec_nonuni}.

For a comprehensive mathematical treatment including continuous and discrete time, please consider, e.g., \citet[][]{Jarrow:15}. The above definition can be also extended for assets with finite maturity, by setting the cash flows to zero after some time. 

A bubble or bubble component $\big(\beta_t\big)_t$ can now be defined as the deviation of market prices $p_t$ from fundamental values, i.e.\ 
\begin{equation}\label{eq:bubble-def-1}
\beta_t = p_t - f_t,
\end{equation}
see for example \citet{Jarrow:15}. That means, 
\begin{tcolorbox}
a bubble is any deviation of a security's price from its fundamental value.
\end{tcolorbox}

The question whether the bubble process $\beta_t$ in Equation~\ref{eq:bubble-def-1} is different from zero, i.e., whether a ``bubbles exists" essentially corresponds to the question whether the efficient market hypothesis holds, i.e., whether market prices are equal to the present value of expected future cash flows, see Definition~\ref{eq:definition_FV}. In this context, it is worth mentioning the debate between \citet{Fama:14} and \citet{Shiller:14} who centered their Nobel Prize lectures around this question.

\paragraph*{Definition in financial mathematics:}  
In general, in models from financial mathematics such as those of \citet{Jarrow:15, Jarrow:16} or \citet{Protter:16}, the conditional expectation in the definition of the fundamental value is defined by use of the unique equivalent local martingale measure (ELMM) for complete markets. There, ``local'' is only needed for continuous time models, in discrete time an equivalent martingale measure (EMM) is the adequate notion \citep[see][]{Jarrow:15}. In theoretical models, the ELMM reflects all current beliefs of the market participants since all beliefs are reflected in the price which is used for valuating the probabilities. In their work, the fundamental value $f_t$ of the risky asset in continuous time is defined via 
\begin{equation}\label{eq:definition_FV_math}
f_t = \mathbb{E}_\mathbb{Q}\left[\frac{p_T}{B_T} + \int_t^T\frac{1}{B_k} \mathrm{d}d_k\ \middle|\ \mathcal{F}_t\right]B_t,
\end{equation}
where $\mathbb{E}_\mathbb{Q}$ denotes the expectation under the ELMM $\mathbb{Q}$, $p_T$ is the liquidation value of the risky asset at time $T$, the process $(B_t)_{t\geq 0}$ reflects the value of the (locally) riskless asset that is used for discounting future cash flows, i.e., $q_k=\frac{1}{B_k}$, and $(d_t)_{t\geq 0}$ is the process of dividend payments of the risky asset. 
Also here, bubbles are defined as deviations of the discounted market price from the fundamental value.

\paragraph*{Continuous time vs. discrete time:}
We also note that things become more complicated in continuous time than in discrete time, e.g., there are different notions of market efficiency in the literature (no arbitrage, no free lunch with vanishing risk, no feasible free lunch with vanishing risk, no relative arbitrage, no unbounded profit with bounded risk) and accordingly different equivalent conditions (equivalent martingale measure, equivalent local martingale measure, strict martingale density, strict local martingale density). In these setting it might happen---if markets are incomplete---that the probability measure used for pricing is not unique, which complicates the notion of bubbles further. For this branch of literature, see \citet{Delbaen:94,Delbaen:98,Cox:05} as well as \citet{Sin:96,Criens:20a,Werner:97,Jarrow:10,Zanten:08,Delbaen:04,Criens:20b,Zeng:09,Platen:20,Platen:21,Guasoni:10,Downarowicz:10}. Especially, work ``on the birth of bubbles'' is notable, too: \citet{Biagini:14,Biagini:15}.

\paragraph*{Infinite vs.\ finite horizon:} 
Another point to discuss is the time horizon. In a literature branch of continuous time models of financial bubbles, for example in the work of \citet{Loewenstein:00} and \citet{Cox:05}, a finite time horizon $T$ for holding the asset is used. This creates new questions, since in common discrete time, rational expectations and rational behavior settings, this could mean that bubbles might be ruled out \citep[][p.7]{Blanchard:82}. More recent work on continuous time models of financial bubbles was conducted by \citet{Jarrow:16} or \citet{Protter:16}. Furthermore, if individuals buy an asset with the purpose of reselling it, they always profit from a price rise relative to the bond. Thus, if they have reasons to believe that demand for the asset will rise, their expectation for the asset price at the liquidation time $T$ rises as well. In this case, it becomes difficult to distinguish fundamental trading from speculation (i.e.\ from buying assets with the purpose of profiting from a higher selling price). Here it is worth noting that there is a generalization of finite and infinite horizons resp.\ a mixture of it: so-called stopping times. These are random times (that are measurable concerning the respective $\sigma$-Algebra) that are stochastic and may be  finite or infinite. Finite horizon models are especially used when financial products with known and finite maturity are analyzed, e.g., options.

\paragraph*{Cash flows: }
Obviously, in real markets (with infinite horizon) it is not meaningful to deal with sums of undiscounted cash flows, since they can be infinite resp.\ they will. The situation is different in settings with a very limited time line, where discounting may be neglected. For example, in economic experiments. There, the distinction between fundamental values as sums of undiscounted expected payoffs (which may be known to the agents) and completely exogenously given (and possibly meaningless) fundamental values is rather fuzzy, hence we refer to Section~\ref{sec_experiment} for this issue.

\paragraph{Present value of ``future'' actual cash flows}
\label{sec_Actual}
Also \citet{Siegel:03} starts by distinguishing between definitions of assets price bubbles that rely on past price data, namely price paths with an notable upwards trend and a sudden crash \citep[cf.][]{Kindleberger:78} and definitions that compare the price with its fundamental, which might be defined as the long-run price in a general equilibrium. Then, \citet{Siegel:03} analyses historic price data and defines a bubble as a price that is beyond the sum of the realized discounted dividends from a retrospective view. However, since, as long as the firm exists, one never knows all dividends, \citet{Siegel:03} proposes to use the (time-weighted) duration, which is in his examples around $27$ years. 

When looking back at historical data, there are actual cash flows that can be used to define a fundamental value. In this case, no expectations are formed, but the actual discounted dividend payments are used to define the fundamental value~$f$. Formally, 
\[f_t =\sum_{k = 0}^{\infty} q_{t+k,t}d_{t+k}, \quad t = 0,1,2,\ldots \] where $q_{t+k,t}$ is a discount factor of the cash flow from time step~$t+k$ to time step~$t$ (which can be derived by bond prices) and $(d_t)_t$ is the deterministic stream of dividend payments \citep[cf.][]{Barlevy:07}. When using actual (historic) cash flows, one does not have to struggle with ambiguous (subjective) probability measures. 

\begin{tcolorbox}Retrospectively, a bubble is a price which lies above the sum of discounted actual pay-offs. However, the time-span has to be ``long enough.''\end{tcolorbox}

\paragraph{Utilities not only from cash flows}
\label{sec:utility_other_than_cash}

Utilities of many assets such as commodities (e.g., gold, currencies, food) or real estate are not (only) derived from expected future cash flows such as dividend payments, as in our explanations above.
When interpreting ``dividend payments'' not only as monetary cash flows, but also considering the monetary valuation of utilities from using the good, the above framework may be adapted to non-monetary payoffs. Such valuations can for example be conducted via assessing the use value (cf.\ ``revealed preferences'') or opportunity costs for equal alternatives for which a direct monetary valuation exists. 
Such valuations  are applicable to subjective, individual, or idiosyncratic utilities, leading to heterogeneous fundamentals, as it is, e.g., important for  investments in ethical funds or religious banks \citep[see, e.g.,][]{Wilson:97,Friede:15}. This way of dealing with non-monetary payoffs is, for example, referred to by \citet[][p.\ 2]{Blanchard:82} when they state that the ``dividends'' in their formula for the fundamental value \begin{quote}`[\dots] may take, depending on the asset, pecuniary or non-pecuniary forms.'\end{quote} 

The valuation of utility other than from future dividend payments can, e.g., be done via prices for new items and second-hand prices \citep{Adland:06}. An important application area of utility valuations is in real estate resp.\ housing markets \citep[cf.][]{Allen:13}. If a house is bought in order to be rented, these rents are future cash flows. However, if the buyer is living in this house himself, there are no direct cash flows. However, the fundamental value can be derived from the rent one would have to pay for this house \citep[see][]{Allen:13}, cf.\ the discussion in Section~\ref{sec_HOUSE}, although these rents might not be observable due to missing comparative material.

In this thought, subjective beliefs about the future become even more important, e.g., when assigning a value to owning an essential good (such that the risk of not having it in the future is minimized). Similar ideas apply to the valuation of gold and currencies\footnote{The functions of money may be defined as a) means of exchange/payment b) measure/unit of value c) means storage d) means of transfer.} and also to fiat money\footnote{Cf.\ the seminal paper of \citet[][]{Townsend:80}, the literature on ``money in the utility function'', and discussions on ``money is trust-based''.} and crypto currencies. See also \citet{Diba:88}. With regard to gold, several utilities may be utilized to calculate fundamentals \citep[see][]{Bialkowski:15}.

Thus, in this sense an asset price is 
\begin{tcolorbox}a bubble, if it deviates from the fundamental value that is computed using a monetary valuation of the utility of using the asset in the present and future.\end{tcolorbox}

\paragraph{Book, resale, and liquidation value as well as Tobin's Q}\label{book}
Another approach to define bubbles is to compare a company's value, which might be book value, resale (i.e.\ liquidation) value, or replacement costs, with its market value, since this shows whether a share has a higher or lower worth than its price \citep{Investopedia1,Investopedia2}, which might be utilized in the so-called `fundamental analysis.' In more detail, the book value is defined as equity in their balance sheets, that is, the difference between its assets and its liabilities. Then, following this definitional category, the fundamental value of one stock share is the value of the underlying firm's equity divided by the number of the issued shares. To provide another example, \citet[][p.\ 1342]{Jarrow:12b} state that, in the context of their bubble model of a stock market,
\begin{quote} ``[\dots] [the fundamental value] $F_t$ represents the liquidation value of the stock if the firm is liquidated at time $t$.''
\end{quote}

A further way to incorporate information available in the balance sheets or books of a firm is to utilize Tobin's Q \citep{Tobin:69}. \citet[][p.\ 217]{Hayashi:82} defines the `marginal~q' as the ratio of ``the present discounted value of additional future (after-tax) profits that are due to one additional unit of current investment'' and the (replacement) price of one investment good; and the `average~q' as ``the ratio of the market value of existing capital to its replacement cost.'' \citet{Hayashi:82} shows that this marginal~q is connected via so-called installation functions with the capital stock of the (investing) firm and its optimal investment (in some investment good) and some other variables. \citet{Miao:15} pick up this marginal q and define---for some (non-obvious) reason---
\begin{tcolorbox}
the fundamental value as the product of Tobin's q (presumably: Tobin's marginal q) and the capital stock of the underlying firm.
\end{tcolorbox}
\citet{Martin:18} connect bubbles to the capital stock.

\paragraph{Market equilibrium}\label{MarketEq}
Besides determining a fundamental value as in the ways discussed above, 
one can also define a fundamental value directly as microeconomic equilibrium of demand and supply factors in the absence of speculation or similar forces.
More specifically, this category is about papers in which the authors distinguish a (theoretical) equilibrium arising equating microeconomic demand and supply functions from an (observed) market price where the latter then includes additional behaviour such as speculation or excessive risk taking. The specific behaviour driving the bubble can but not necessarily has to be explicitly specified in the respective publication. Here, we understand speculation as the action of buying or selling an asset for the only purpose of earning from capital gains \citep[cf.][]{Juvenal:15}. Excessive risk taking is defined as actions in which traders take on higher risk than would be explicable by rational behaviour \citep[cf.][]{Acharya:12}. For more detailed explanations of the concept of a competitive market equilibrium, we refer to microeconomic textbooks \citep[cf.][]{Mas-colell:95, Elsner:14}.

We furthermore extend the category of fundamental values as \textit{market equilibrium} to econometric approaches in which the authors use empirical data on supply and demand side aspects to determine a fundamental value of the asset, and then compare it with its market price. For example \citet{Zhang:15} consider empirical data on production of crude oil, imports, inventories and an industrial production index to determine a fundamental value by employing a linear regression model. The bubble is then defined as the difference between this fundamental value and the observed market price.

Overall, we point out that the market equilibrium category is applicable to both, commodities without regular dividend-like payments (such as gold or crude oil) as well as financial assets with regular dividend payments, for which a definition with a net present value of future cash flows would also be plausible \citep[see e.g.][]{Barberis:18a}.
With regard to this, we acknowledge that depending on the assumptions made (e.g. regarding rationality of market participants) a fundamental value defined via expected future cash flows may be equivalent to the one defined via supply-demand equilibria. However, since under different assumptions such equivalence might no longer hold and since we are interested in the definitions that are practically used in scientific publications, we consider these concepts separately.

To sum up, this category subsumes definitions of fundamental values in terms of a supply-demand equilibrium without speculation (or a similar mechanism alternatively). A bubble is then \begin{tcolorbox}a deviation between the (observed) market price and the (theoretical) demand-supply equilibrium.
\end{tcolorbox}

\paragraph{Heuristics}\label{Heu}
Furthermore, a fundamental value can be defined or approximated 
by using simplified rules (=heuristics) such as asset price averages, or different types of moving averages of the asset price. In this setting, a bubble is then determined by  
\begin{tcolorbox}the deviation of the market price from its long run price average or other heuristical determination of the fundamental value.\end{tcolorbox}
For example \citet{Baur:14} approximate the fundamental value of gold by an exponential moving average. 

The heuristics considered in this category may rely only on the asset price itself for fundamental value computation but may also take into account further information about the asset itself or the surrounding market conditions. Thus, we also include econometric estimations of macroeconomic properties that do not explicitly refer to demand and supply factors. For example, a model for determining the fundamentals of house prices could include land prices, income per capita, population and interest rates, as for example used by \citet{Dreger:13}, see also \citet[][]{Ambrose:13} for a similar example. We remark that papers using econometric approaches to determine fundamental values might be in either or both, the category ``heuristics" as well as the category ``market equilibrium", depending on how concretely supply and demand functions are described. 
For econometric approaches, a bubble component of an asset price then arises from
\begin{tcolorbox}the
the difference between the market prices and the econometric prediction.\end{tcolorbox}

\paragraph{Exogenously given}\label{sec_exo}
Although, as described above, one of the main problems when defining bubbles as deviations from fundamentals is that it is rather unclear what the true fundamental value is (if there is one) and that such a value is difficult or impossible to calculate, there are some research fields where fundamentals are known. One possibility is that fundamentals are exogenously given and ``meaningless,'' i.e., from the point of view of a trader it is either assumed that possible dividends are too small to be taken into account compared to possible gains from price changes or there are questions like ``If all traders believed that the fundamental value is $f$ given, under a specific trader type combination what happens?'' are investigated, which lies in the field of heterogeneous agent models (HAMs), Section~\ref{sec_HAMS}. Such exogenously given fundamentals can be constant, deterministic functions, or even stochastic processes. Also when testing for bubbles (see Section~\ref{sec_tests}), abstract fundamentals without meaning can be assumed. Or when dealing with gold, constant fundamentals are sometimes utilized \citep[cf.][]{Baur:14}.

We note that another possibility for how fundamentals can be known is in experimental economics. The researcher who designs the experiment specifies the fundamentals in advance and observes how the traders, i.e.\ the agents, act---sometimes they know the fundamentals, too, sometimes not. Thus, all agents in an experiment may know future cash flows and, if they are uncertain, the respective probabilities. But this is not what we define as exogenously given since, as explained, the experiment designer and/or the agents use future expected cash flows. Hence, such papers are categorized in the respective future cash flow area.

\subsubsection{(Non-)uniqueness of fundamentals}\label{sec_nonuni}
In the above definitions, it was not specified which beliefs or probability measures were used to arrive at the respective expectations.  
For example, in one of the bubble definitions used by
 \citet{Barlevy:07} individual probability measures for defining individual fundamental values are considered, 
 i.e., each investor has her own beliefs about her future and can thus calculate her own fundamental value. In this case \citet{Barlevy:07} defines (in Box~1 on Page~47) overvalued assets, i.e.\ bubbles, as follows:
\begin{quote}
`However, as Harrison and Kreps (1978) and Allen, Morris, and
Postlewaite (1993) point out, it is still possible to talk about an asset as unambiguously overvalued if it exceeds the fundamental value \textit{any} trader in the market would assign to it---that is, if there is no trader who would be willing to buy the asset at the going price if he had to hold the asset forever and never sell it.'
\end{quote}
To sum up, 
\begin{tcolorbox}
if traders have various/subjective/idiosyncratic beliefs/expectations for the fundamentals, a price is a bubble if it is above their maximum. 
\end{tcolorbox} Analogously, this definition applies for negative bubbles.
Another source of heterogeneity in the definition of fundamental values is the choice of probability measure for the computation of expectations. Definitions of bubbles that depend on the probability measure being used can be found in the mathematical finance literature, e.g., \citet{Schatz:20}. \citet{Schatz:20} consider $\mathbb{Q}$-bubbles which denote the difference between the market price and a fundamental value computed using the probability measure $\mathbb{Q}$. 

While some economical papers use---as just mentioned---the underlying probability measure to describe the (rational) believes of the traders about the future without stating stochastic details, it is a key ingredient in mathematical finance: for the purpose of pricing, the/an equivalent (local) martingale measure is used while for the purpose of risk quantification the so-called market measure is consulted.

\subsubsection{Asset price properties}\label{sec_def_price}

Another way how bubbles may be defined is via price properties. This can be done additionally to properties relying on fundamentals or instead of them. Furthermore, the definitions can be described via words or via formulas. For example, one may define a bubble as 
\begin{quote}`a sharp increase in an asset's price followed
by a steep decline,' 
\end{quote}see \citet[][p.101-102]{Barberis:18b}\footnote{We note that \citet{Barberis:18b} does not define a bubble that way but states that this is `the most essential feature of a bubble'.}, or in formula via price paths that show a hyperexponential growth. We note that papers that, e.g., mention in the introduction that bubbles are characterized by high volatilities, but use only fundamental properties in their analyses, in the border case we do not categorize them as both fundamental and price properties but only as via fundamentals. 

\paragraph{Boom}\label{par_boom}

While sometimes \citep[e.g.,][]{Kindleberger:05} bubbles are defined verbally via price rises, others use such images only for motivation or interpretation of results (cf.\ Section~\ref{sec_int}). When price properties, esp.\ booms or rising prices are used, these properties are often expressed in mathematical terms. For example, \citet{Girdzijauskas:09} start their paper with \begin{quote}`A stock market bubble in the financial markets is the
term that is applied to a self-propagating rise or increase
in the share prices of stocks in a particular industry
or sector. The term ``stock market bubble'' can
only be used with any certainty in retrospect when
share prices have since fallen drastically or crashed.'\end{quote}However, in their analysis, they translate these verbal definitions into formulae using elasticities or sensitivities of return rates. While some papers, e.g., \citet{Fry:16}, explicitly connect definition including fundamentals and, in this case: tests, excluding fundamentals (and utilize price path properties like booms), others focus very much on price properties, e.g., \citet[][p.\ 1645]{Yan:10}, on \begin{quote}`faster-than-exponential growth'\end{quote} of the price. Also \citet[][p.\ 115]{Sornette:14}, although recognizing fundamentals, define bubbles \begin{quote}`as the ``super-exponentially'' accelerating rise of a price [\dots].'\end{quote} For a connection of fundamentals and price rises, confer also \citet{ZhangYao:16}.

\paragraph{Crash} \label{sec_crash}
There is work, most prominently by Anders Johansen or Didier Sornette and co-authors \citep[e.g.,][]{Jiang:10}, which is explicitly interested in crashes. There, bubbles can, but do not have to, end---after the so-called critical time---in crashes, thus, the possibilities for crashes are used to (indirectly) define bubbles. Also \citet[][p.\ 163]{Johansen:03} is focused on crashes, which are consequences of bubbles that are connected to \begin{quote}`log-periodic power law precursors.'\end{quote} These so-called LPPL models can be used to test for bubbles and crashes, see Section~\ref{sec_tests} \citep[and, e.g.,][]{Bree:13}.

However, there are also authors who are not so much interested in the crash itself but mostly in the bubble. They characterize---possibly only verbally---bubbles by the crash that comes eventually or that is at least possible. In this sense \citet[][p.\ 203]{Barberis:18a} write:
\begin{quote}`The bubble eventually ends with a crash, in which prices collapse even more quickly than they rose.'\end{quote} Such ideas one can find also in the papers of \citep{Barlevy:07,Allen:13}; e.g., when the latter ones call ``boom-bust cycles.''

When defining bubbles via crashes, there is an obvious question that arises: What is a crash? Hence, the problem of defining bubbles is just transferred to the problem of defining crashes. This can be done, e.g., via percentage drawdown, as conducted by \citet{Greenwood:19}, i.e., a crash is an event when a stock loses in a time interval with predefined span a predefined ratio of its value---or via connections to \begin{quote}`phase transitions'\end{quote} as in physics \citep[][p.\ 493]{Kaizoji:00}. Thus, if a crash or a possibility of a crash is considered a defining feature of a bubble, articles are categorized into this category.

\paragraph{Volatility}\label{par_vola}
Last, there is a connection of bubbles to volatility---beyond the obvious one that under the assumption of not too volatile, or loosely spoken: not to wild, fundamentals, all highly volatile prices must rather often deviate from their fundamentals, which may be used directly to define bubbles (together with other properties) as done by \citet{Glaeser:15}. For example, in \citet{Fry:16}, although bubbles are characterized via prices increases and deviations from intrinsic values---which cause a fall afterwards---it is under specific assumptions shown that this can be tested for via anomalies in growth rates and, which is important, volatilities. 

\subsubsection{Extensional definition}\label{sec_def_expl}
One can also define the term bubble by explicitly listing all empirical phenomena that are said to be bubbles. For example, in \citet{Sornette:18} 40 historical/empirical price/market paths are defined as bubbles. But note that the authors do not say that this list is exhaustive. 

\subsubsection{Other}\label{other}
In the category ``other", we collect characterizations of bubbles that do not fit to any other category but are too rare to get their own one. These might be either completely different, or somewhat similar but not equivalent to the categories before.
For example, \citet{Clark:10} compare growth rates of the income of households and a time series of housing prices around the time of the 2008 financial crisis, applying econometric methods to the UK housing market. A bubble exists in the time series, if, in the time period of interest, housing prices grow significantly more than household income. The key difference of this approach to the definitions above is their comparison of growth rates instead of absolute values of market prices and fundamental values.

\subsection{Fields of application}\label{sec_appl}
In this subsection, we present some application fields for bubble definitions. Furtheron, we link them to the definitions themselves. The reader may note that different definitions often have different applications and that some of those may be used for real-time bubble detection while others may not.

\subsubsection{Bubble tests}\label{sec_tests}

Empirical tests for bubbles aim at identifying bubbles from empirical data on asset prices, sometimes with and sometimes without additionally considering data on fundamentals. In most tests, bubbles are understood as deviations from the sum of future expected and discounted cash flows (Definition category \ref{par_cashflow}). However, the operationalizations differ widely, often relying on price path properties only, e.g. volatility or growth, as in the definition categories in Paragraphs \ref{par_vola} and \ref{par_boom} or on heuristics (Paragraph \ref{Heu}). Examples of heuristics in bubble definitions are ``a period with a nonzero median in excess returns" \citep[][p.\ 621]{Evans:86} or price deviations of more than one (or two) standard deviations from a moving average of the price process \citep[][]{Vogel:18}. In the following, we discuss three test methodologies commonly found in recent bubble test literature. For a review on earlier bubble tests, see \citet{Camerer:89}.

A large class of bubble tests relies on the concept of rational expectation bubbles (or rational bubbles, growing bubbles, see \citet{Gurkaynak:08}). Rational bubbles are deviations from fundamental values that arise as the infinitely many (non-fundamental) solutions to the pricing Equation \eqref{eq:price-equation} for which the so-called ``transversality condition" (also: ``no-bubble condition'') is not fulfilled (cf.\ Section~\ref{sec_def_fund}, \citet{Blanchard:82}). Such rational bubbles are characterized by exponentially growing price paths which can be detected statistically. This in turn is the basis for the use of augmented Dickey-Fuller (ADF) tests for bubbles \citep{PhillipsWuYu:11, PhillipsYu:11, Phillips:15}, which identify whether asset price data or the price-dividend ratio follow a unit root or an explosive root process \citep[][p.\ 206]{PhillipsWuYu:11}, where the latter is an indicator for a bubble. If only price data is analyzed, this operationalization falls into the category of characterizing bubbles as price boom, cf. Section \ref{par_boom}, otherwise we categorize it into ``deviation from fundamentals", cf. Section~\ref{par_cashflow}.

Another methodology of real-time bubble detection are log-periodic-power-law (LPPL) tests which are based on detecting a hyperexponential price path growth, i.e. asset prices exhibiting faster than exponential growth. While originally derived from a definition of bubbles as deviation from fundamentals \citet{Zhang:16}, the test methodology does not require any data on fundamentals itself, but relies on growth properties of price data. The LPPL methodology is based on the idea that stock market dynamics can be modelled as a self-organised system exhibiting critical behaviour, similar to models of earthquakes \citep[cf.][]{Sornette:96}. Bubbles are modeled as time periods in which the logarithm of the asset price is following a power law with accelerating periodic fluctuations leading up to a critical time $t_c$, at which, at the latest, the bubble collapses \citep{Johansen:00, Sornette:03}. While the actual collapse time in the model is random, the critical time $t_c$  or time window of the bubble can be estimated, which is found to be a useful tool, for example, by \citet{Cajueiro:09} in the case of Brazilian stock market data, \citet{Jiang:10} for Chinese stock market data, and \citet{ZhangYao:16} for oil price data.

In the financial mathematics literature \citep[see][]{Protter:16,Jarrow:16} bubbles can be characterized as price processes with sufficiently high volatility. They start from a model of the asset's market price using a stochastic differential equation including a volatility function \citep[][eq.\ 6]{Jarrow:16}. As in the previous paragraphs, a process is defined to be a bubble process if it deviates from the fundamental value. For a certain subclass of such bubbles, the authors state a condition for this process to be a bubble that solely depends on the volatility function of the process \citep[][eq.\ 8]{Jarrow:16}. In this characterization, the fundamental value is not present. If the volatility fulfills the respective condition, the process is said to be a bubble process, which is used in empirical tests for real-time bubble detection \citep{Jarrow:16}. 
There are further equivalent properties for bubble processes \citep[see][]{Jarrow:12a}. Based
on this, further tests are constructed, which have the advantage that much less assumptions on the underlying structures have to be made than in traditional tests \citep{Jarrow:16}. 

To sum up, one can say: It can be possible to define bubbles via
deviations from fundamentals, but test them only by use of price properties,
e.g., explosive root processes, hyper-exponential growth, or certain volatility properties which in turn allows for bubble detection in real-time.

\subsubsection{Heterogeneous agent models}\label{sec_HAMS}
In some heterogenous agent models (HAM), the fundamental value is given from an external source (Section~\ref{sec_exo}) \citep[see, e.g.,][]{Horst:05, Harras:11, Thurner:12}. These exogenously given fundamental values can be common to all traders or individually different, as in \citet{Youssefmir:98}. The idea behind that is the aim to analyze the mutual effects on markets or prices from and on traders. If the effects of this loop is of interest, the meaning of the fundamental (that might be noisy and/or stochastic) is not that important, but only that the traders react to it. A basic approach in HAM, dating back to \citet{Zeeman:74}, \citet{Beja:80} and \citet{Day:90} is to assume that there are two types of traders: chartists and fundamentalists. The former trade according to price trends, the latter based on fundamentals. Research topics are, e.g., whether fundamentalists can stabilize markets or how stylized facts can be rebuild,\ \citep[e.g.,][]{Beja:80,Day:90,Franke:16}. Very readable overviews of this branch of the literature can be found, e.g., in \citet{Hommes:06,Chiarella:09,Hommes:21}. On the other hand, there are papers in the heterogenous agent literature in which the fundamental value is defined as the present value of future expected dividend payments, i.e.\ a price that solves the Euler-Equation (\ref{eq:price-equation}) \citep[][p.\ 171]{Hommes:21}. Furthermore, there are instances in which no absolute fundamental value but rather exogenous positive or negative ``fundamental signals'' are provided, which can be common to all fundamental traders \citep[see, e.g.,][]{Hott:09} or individual \citep[see, e.g.,][]{Biondo:13}. Moreover, the fundamental value can be approximated by an exponentially moving average of past prices \citep[see, e.g.,][]{Baur:14}.

\subsubsection{Experimental economics}
\label{sec_experiment}
As explained above, when defining bubbles with respect to some fundamental value, usually the difficulty arises how fundamentals shall be measured. One exception is the field of experimental economics. In this field, the behavior of individuals in experimental economic settings is investigated. Under tight and observable conditions it is for example analyzed under which constraints agents are willing to pay ``too much'' for an asset, that is, when they produce a bubble. One may consider \citet{Palan:13} for a review on this topic. The advantage of this research field is that the fundamentals such as dividend payments and their probabilities are exogenously given and known to the experimenter---and depending on the experimental setting also to the agents. There, a price path is called a bubble if the price deviates from its fundamental \citep[see, e.g.][]{Cheung:12, Hirota:16, Lei:01}. 
It is also important to note, that in the framework for experimental asset markets reviewed by \citet{Palan:13}, future cash flows are not discounted in the calculation of the intrinsic value. 

Questions in experimental markets go beyond the Boolean of whether there is a bubble or not. Also the size or strength of bubbles is measured, which in converse can be seen as additional aspects of bubble definitions. Following \citet{Stoeckl:10} any bubble measure should (i) relate the price and the fundamental value, (ii) be monotone in the difference between the price and the fundamental value, (iii) be independent from the absolute level of the fundamental value, and (iv) be independent from the time horizon \citep[see also][]{Cheung:12,Haruvy:07,Noussair:01}.

\subsubsection{Housing and real estate prices}
\label{sec_HOUSE}
Housing markets are different from other financial asset markets, due to its individual and dispersed nature \citep{Glaeser:15}. In the review on housing bubbles by \citet{Glaeser:15}, many such peculiarities are pointed out, for example the wider ownership of housing, which makes it more a consumption good rather than a ``usual'' financial asset. Also, due to the decentralized transactions, non-comparable assets, and higher search costs, there is much less of a single housing price, compared to, e.g., stock shares traded on a central exchange. This makes using the ``standard tools" from financial asset pricing that rely on the existence of a (unique) fundamental value much more difficult to apply. Nevertheless, there are several empirical approaches to determine whether certain housing prices are in a bubble or not. We distinguish three main approaches: First, estimating fundamentals using the price-rent ratio \citep[e.g.,][]{Ambrose:13}, second, using current macroeconomic variables to estimate changes in fundamentals \citep[e.g.,][]{Bourassa:01}, and, third, relying on price path properties only, following the method by \citet{Phillips:12} \citep[e.g.,][]{Yiu:13}. For details on the latter, see also Section~\ref{sec_tests}.

In an econometric analysis, \citet{Ambrose:13} conduct an 
autoregression analysis i.a.\ with 
housing prices and rents. 
While the term ``bubble" is not explicitly defined, the authors emphasize the periods in which the actual price-rent ratio deviates from the theoretical value.
The method is based on \citet{Campbell:09}, who do not explicitly use the term ``bubble,'' but analyse \begin{quote}`fundamental sources of variation in rent-price ratios'\end{quote} \citep[][p.\ 101]{Campbell:09}.

\citet[][p.\ 3]{Bourassa:01} identify the value of houses as the \begin{quote}`sum of the structure and land values.'\end{quote} This ``value'' can be identified with what we call fundamental value. In their data analysis, they estimate the change in fundamental value of housing using a linear relationship with 
macroeconomic variables. 
Furthermore, using insights from bubble formation, \citet{Bourassa:01} build an estimation model for the bubble term in house prices.
However, there are also examples of unique fundamentals in housing models: \citet[][p.\ 40]{Allen:13} propose  
model of the housing market in which a unique fundamental price can be understood as \begin{quote}`expected flow of housing services.'\end{quote}

Here, we want to add an interesting consideration to think about: If there is a bubble in housing markets, rents will likely go up, too. But if rents are high, high prices for houses are no bubbles, where the bubble definition relies on future cash flows. Thus, esp.\ in the housing markets, there might be a circular definition.

\subsubsection{History of economics}
\label{sec_history}
At this point, we note that there is another discipline in economics that deals with bubbles (and, especially, with crashes), namely history of economics. There are the prominent books of \citet{Aliber:15} and of \citet{Garber:01}. Often in history of economics, bubbles are defined in an exemplary way: that is, it is said that, e.g., the Tulipmania, the dot-com bubble, the Mississippi bubble were ``bubbles'' (see Section~\ref{sec_def_expl}). Then it is explained how such a bubble could occur or what type of bubble it was.

\subsubsection{Machine learning}
The definition by using examples of Section~\ref{sec_def_expl} is really interesting because it provides a completely new way of bubble detection. With the help of AI or machine learning,\footnote{The authors are grateful to Marina Eliza Spaliara (Adam Smith business school, University of Glasgow, Glasgow, Scotland, UK) for suggesting to incorporate the connection between economic bubbles and machine learning.} one can try to find bubbles if one trains a neural network. See \citet{Ozgur:21} and \citet{Basoglue:21} for these topics. Perhaps the bubble phenomenon is high-dimensional (fundamental values, prices, supply, demand, politics, etc.) and can be reduced in dimensionality with the help of AI. However, one has to keep in mind that rare events are hard to predict. 

\subsubsection{Policies}
A task in the sciences is to identify bubbles on the fly---and closely related to this is the question of what implication the issue of bubbles has for policymakers. \citet{Barlevy:07} and \citet{Cogley:99} ask the question whether it is desirable from the point of view of a society to let asset price bubbles burst via policies \citep[cf.][]{Allen:13}. For this reason, \citet{Barlevy:07} first distinguishes between the term bubble as used in the popular press and the term bubble as ``most economists'' would define it. There, the first term can only be used in a retrospective way, i.e., a bubble is a bubble because it crashed. The second definition uses the fundamental value of the asset. Although such a bubble does not have to crash certainly, there is no reason why it should not crash. \citet{Barlevy:07} brings it to the point by saying that an asset's price is called a bubble because it could crash. Moreover, he notes that there are also reasonable cases for prices to rise sharply and then fall very quickly that are not bubbles, namely when an innovation is traded the price is high until competitors imitate the innovation \citep[cf.][]{Camerer:89}. This rewards the innovation and increases the incentive for competitors to be fast. 

\citet{Barlevy:07} finds (building i.a.\ upon the studies of \citet{Milgrom:82}, \citet{Tirole:82}, \citet{Townsend:80}, and \citet{DeLong:90}) that there are only special circumstances under which bubbles can occur. One of those possibilities is that there is already some form of inefficiency in the market.\footnote{The other possibilities arise when one assumes infinitely many traders or if there are differences in the initial belief or beliefs about the existence of irrational traders.}
When one wants to let a bubble burst only when society as a whole is better off as a result and no one is worse off, it is highly questionable whether policy makers should let bubbles burst at all. Indeed, when the market was already inefficient, society may be worse off as a result of the bursting of the bubble \citep{Barlevy:07}. At this stage we note that the definition of bubble used by \citet{Barlevy:07} is reasonable and widely accepted, it is not suitable to detect bubbles in real-time. We note that the argument that a trader does not pay more than the sum of expected discounted future dividends only holds if one assumes a risk-neutral trader---but traders can also be risk-affine. Furthermore, although the probability of a bubble may be very small, the expected losses to the society from the bubble may be very large.

\section{SYSTEMATIC REVIEW}\label{sec_rev}
We conduct a systematic review to assess the frequencies of the different bubble definitions described in Section~\ref{sec_def_struct}.
In this section, we first summarize the methodology (Section~\ref{sec_method}), and then show the results (Section~\ref{sec_results}). In the end, we interpret (Section~\ref{sec_inter}) and discuss (Section~\ref{sec_disc}) these results.
Our results confirm the heterogeneity of beliefs and thoughts of economists on which is the most wide-spread bubble definition, visible in the quotes in Section~\ref{sec_Quotes}.

\subsection{Method}\label{sec_method}
To conduct the systematic review, we used the search engine Scopus, due to its flexibility regarding data export and filtering options.
For an overview of different search engines, their benefits and disadvantages, see \citet{Gusenbauer:20}. Where applicable,  
our research was guided by the reporting guidelines of the Preferred Reporting Items for Systematic reviews and Meta-Analyses (PRISMA) statement for transparency of our procedure \citep{Page:21}.\footnote{The PRISMA statement was developed primarily for syntheses of the results of quantitative studies. Since we synthesize definitions (of bubbles) used in various studies and not their research results, only 14 of the 27 checklist items of the PRISMA statement apply. The reported items in this paper are 1, 3-9, 16, 23, 24b, 25-27. The studies included in the analyzed set can be found in the supplementary material (checklist item 17). The study was not registered and no review protocol was prepared. The creation of a PRISMA flow chart was omitted since it would only contain three entries: 
124 papers in the original set, 123 after removing one retracted paper and 122 after removing off topic papers.}

The keywords used for the search are the following: 
``market bubble'',
``market bubbles'',
``price bubble'',
``price bubbles'',
``asset bubble'',
``asset bubbles'',  
``bubbles and crashes''.
At least one of these keywords must appear in title, abstract or keywords\footnote{`TITLE-ABS-KEY'} of an article. These keywords were found via pre-searches on Google Scholar as well as Scopus. In order to exclude papers that are not directly related to financial bubbles, we further restrict the search to papers in which the keywords of the articles (registered in the search engine Scopus) specifically contain the phrase ``bubble" (except for the keyword ``bubble (in fluids)"\footnote{`EXACTKEYWORD'---``Bubble", ``Asset Bubble", ``Price Bubble", ``Bubbles And Crashes", ``Housing Bubble", ``Stock Market Bubbles", ``Financial Bubbles", ``Financial Bubble", ``Speculative Bubbles", ``Asset Price Bubble", ``Rational Bubbles", ``Housing Bubbles", ``Multiple Bubbles", ``Stock Market Bubble", ``Real Estate Bubble", ``Housing Price Bubble", ``Speculative Bubble", ``House Price Bubbles", ``Market Bubbles", ``Stock Price Bubbles"}). Furthermore, we restrict our search to articles in English with more than 15 citations. This results in 123 articles to be reviewed.\footnote{In our search, we needed to exclude one paper, that has been retracted by the authors, resulting in 123 instead of 124 articles.}  Among these papers, one is unrelated to the topic of financial bubbles which results in a final set of 122 papers on financial bubbles. The data was downloaded from Scopus on December 22nd, 2021 on 4:50pm CET. Here, we note two points. Firstly, the minimum citations number of 15, and secondly, the download date 2021, both, exclude very new papers. However, when assuming that the share of the different bubble definition possibilities used by researchers do not vary  much over time, 
we can conclude that the results regarding fractions of definition types in the literature will be also applicable to newer research publications. For an overview of the numbers of citations at the time of data retrieval as well as the distribution of publication years, see Figure~\ref{fig_hist}.
The full list of articles in our data set can be found in the supplementary material.

\begin{figure}[h]
\centering
\includegraphics[width=\textwidth]{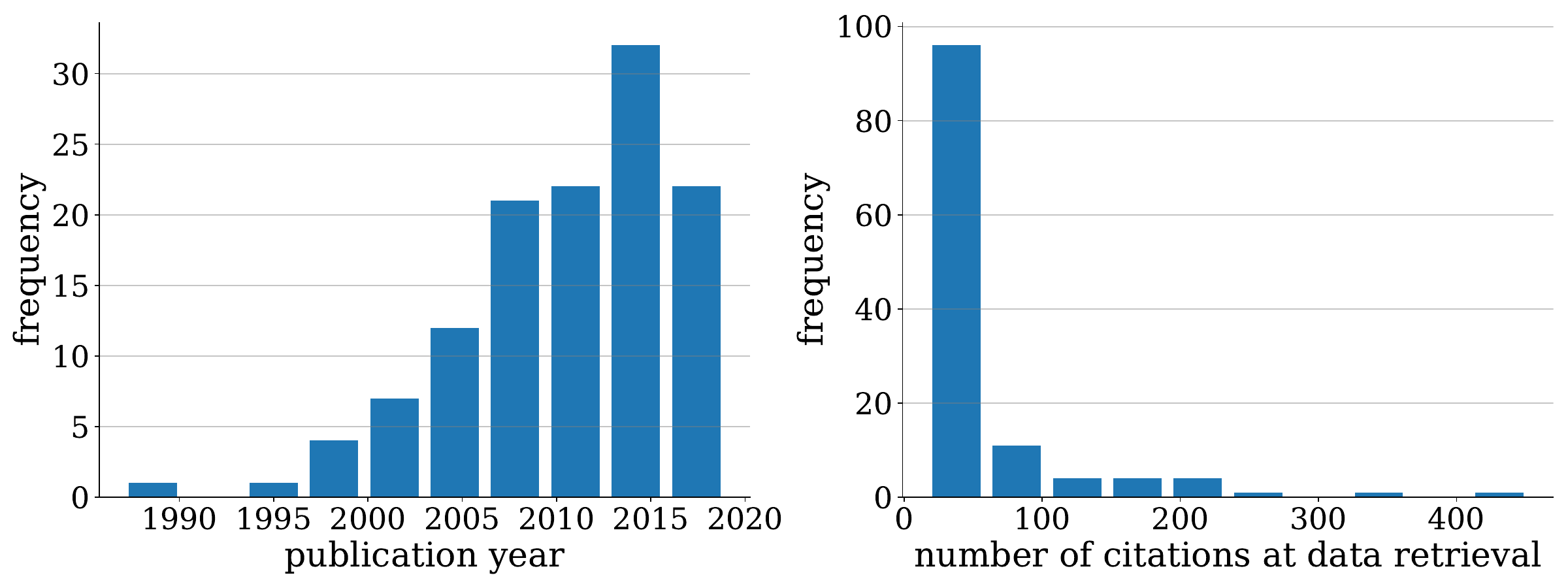}
\caption{Overview of publication years and citations in the analysed literature set of 122 papers, at time of data retrieval (December~22nd,~2021); Graphic generated in Python.}
\label{fig_hist}
\end{figure}

Since the aim of our systematic review is to find quantitative answers on which definitions of bubbles are used, 
we categorize the papers according to the types of definitions used. The definition categories were created after an initial screening of a smaller sample of papers and iteratively updated throughout the review procedure, which resulted in overall 15 final categories related to definitions, see supplementary material. Additionally, we categorize the articles concerning their general field, whether they are theoretical or practical, etc. That means, first we sort the papers regarding their fields: Heterogeneous Agents and Behavioral Finance; 
Macro and General Economics; Econophysics; Experimental Economics; Real Estate and Housing; (Crypto) Currency; Finance and Banking; Policy and Politics; Mathematical Finance; Trade; Commodities, Agriculture, and Energy; and others (incl.\ i.a.\ History, Geography). All categorizations were conducted first by both authors of this paper separately. Then, the deviating items were discussed until a consensus was reached. No automation tools such as machine learning were applied in the categorization process. 

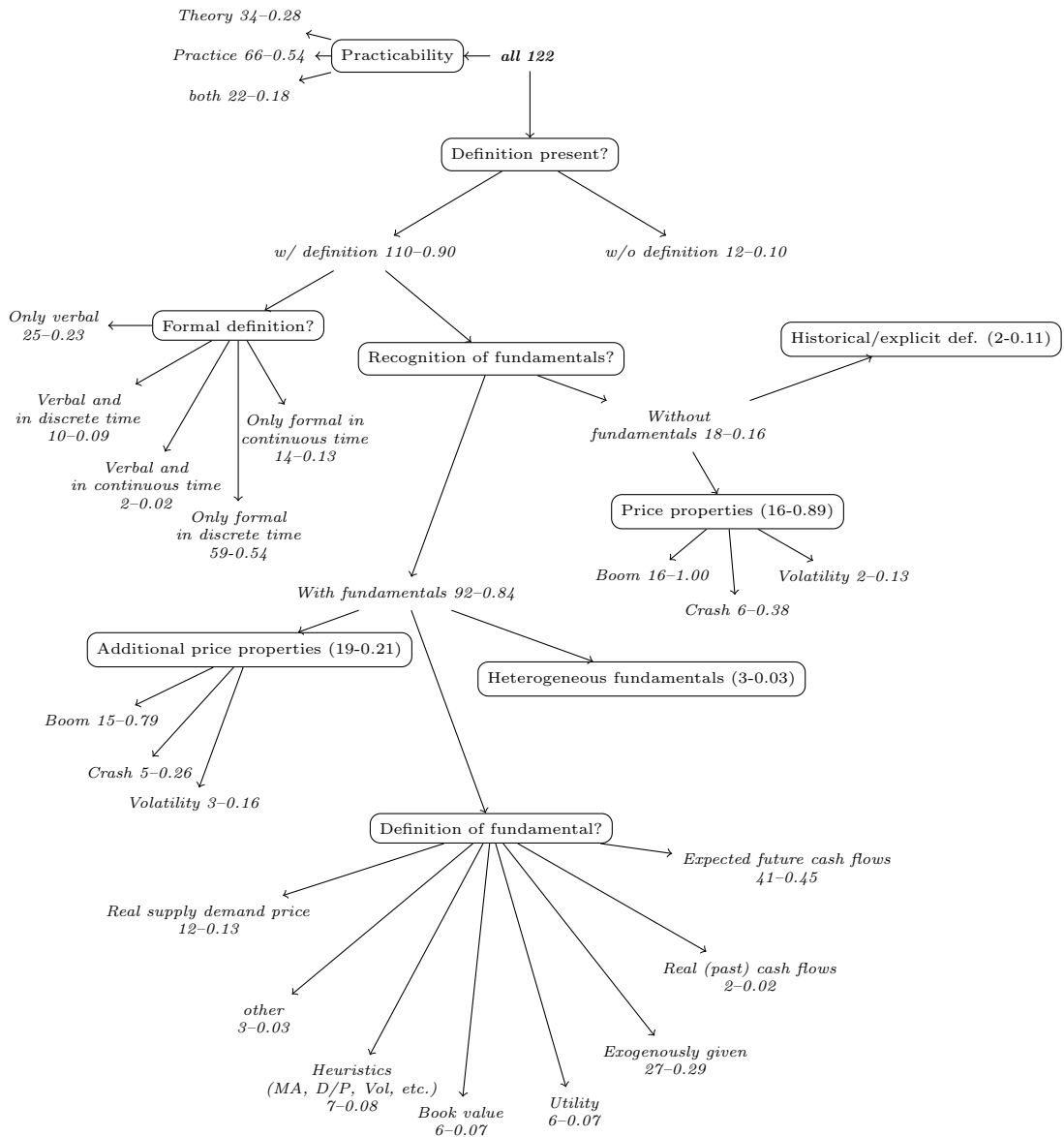
\begin{figure}
\begin{tikzpicture}[q/.style={font=\tiny,rectangle,rounded corners,draw},
a/.style={font={\tiny \it},align=center},edge from parent/.style={draw,->},scale=0.9
]
	\node[a]{\textbf{all 122
}}
		child[grow=180,level distance=2cm]{node[q]{Practicability}
			[level distance=2.4cm,sibling distance=.6cm] child{node[a]{Theory 34--0.28}}
			child{node[a]{Practice 66--0.54}}
			child{node[a]{both 22--0.18}}
		}
		child[grow=270]{ node[q]{Definition present?}
		[sibling distance=5cm]
			child{node[a]{w/ definition 110--0.90}
				child[grow=210,level distance=2.2cm]{node[q]{Formal definition?}
				    [level distance=2.8cm,grow cyclic,counterclockwise from=180,sibling angle=30]
					child{node[a]{Only verbal \\ 25--0.23}}
					child{node[a]{Verbal and \\ in discrete time \\10--0.09}}
					child{node[a]{Verbal and \\ in continuous time \\2--0.02}}
					child[level distance=3.2cm]{node[a]{Only formal \\ in discrete time \\59-0.54}}
					child[level distance=2cm]{node[a]{Only formal in \\ continuous time \\14--0.13}}
				}
				child[grow=320,level distance=2.5cm]{node[q]{Recognition of fundamentals?}
					child[grow=250,level distance=3.8cm]{node[a]{With fundamentals 92--0.84}						
						child[grow=200,level distance=2.5cm]{node[q]{Additional price properties (19-0.21)}
							[grow cyclic,clockwise from=250, sibling angle=22]
							child{node[a]{Volatility 3--0.16}}
							child{node[a]{Crash 5--0.26}}			
							child{node[a]{Boom 15--0.79}}											
						}
						child[grow=290]{node[q]{Definition of fundamental?}
							[grow cyclic,clockwise from=-8,level distance=4.5cm,sibling angle=22]
							child{node[a]{Expected future cash flows \\41--0.45}}
							child{node[a]{Real (past) cash flows \\2--0.02}}
							child{node[a]{Exogenously given \\27--0.29}}
							child{node[a]{Utility \\6--0.07}}
							child{node[a]{Book value \\6--0.07}}
							child{node[a]{Heuristics \\(MA, D/P, Vol, etc.) \\7--0.08}}
							child{node[a]{other \\3--0.03}}
							child{node[a]{Real supply demand price \\12--0.13}}
						}
						child[grow=340]{node[q]
								{Heterogeneous fundamentals (3-0.03)}												
						}
					}
					child[grow=340,level distance=3cm]{node[a]{Without\\ fundamentals 18--0.16}
						child[grow=300, level distance=1.5cm]{node[q]{Price properties (16-0.89)}						
						[grow cyclic,clockwise from=330, sibling angle=55]
							child[level distance=2cm]{node[a]{Volatility 2--0.13}}
							child{node[a]{Crash 6--0.38}}
							child{node[a]{Boom 16--1.00}}
						}
						child{node[q]{Historical/explicit def. (2-0.11)}
						}
					}	
				}
			}
			child{node[a]{w/o definition 12--0.10}} 
		}
;
\end{tikzpicture}
\caption{Results for categorizing the papers in the systematic review\label{tree2}, starting with the set of 122 papers on financial bubbles. The numbers next to the nodes in the format N--0.xx refer to the number of papers N in that node, as well as the fraction 0.xx of papers in this node divided by the number of papers in the parent node. Note, that due to overlapping categories, the sum of fractions in all respective child nodes can be greater than one. Boxed nodes represent questions or topic headlines while in the unboxed nodes answers or categories are provided. However, when there is an N--0.xx in a boxed node, we omitted the answers ``yes'' (to which N--0.xx refers) and ``no''---and when there are again subanswers we additionally omitted the question ``if yes: which one?''.}
\end{figure}

The various bubble definition categories can be found in a tree model in Figure~\ref{tree2}. The category \emph{practicability} refers to whether the results of the paper are only theoretical (sub-category \emph{theory}), or whether the paper uses empirical data or other aspects directly related to applications (sub-category \emph{practice}), or \emph{both}. That means, if the paper has some theoretical contribution, e.g., if it develops a new formal model, it's classified as theory. If there is an empirical contribution, such as an empirical data analysis, or if the paper has a focus on specific policy advice, it is practical.

After determining whether there is a definition present or not, or if the paper is not relevant for the topic of asset price bubbles, we further differentiate the category \emph{with definition} in \emph{discrete} or \emph{continuous time.} We have a look at whether the definition is given by the help of formulae or whether the definition of a bubble is only defined via text (\emph{only verbal}), or \emph{both}. Then, we come to one of the main distinctions, namely whether the definition of bubbles use some form of fundamental value information or not. In the category \emph{with fundamentals}, the bubble is then defined as deviation from these fundamentals. In the category \emph{without fundamentals}, bubbles are explicitly (see Section~\ref{sec_def_expl}) defined or only price path properties (see Section~\ref{sec_def_price}) are considered, which can be properties regarding the \emph{volatility} of the price, or specifically the property that the price rises (\emph{boom}) or price dynamics, that (in addition to the boom dynamics) also require a subsequent \emph{crash} to be identified as a bubble. 

Going further through the tree of Figure~\ref{tree2}, we consider three
aspects of definitions that rely on fundamentals of bubbles. First, whether the definition of the fundamental value is the same for all traders, i.e. \emph{unique}, or whether traders can have their individual fundamental value which then also leads to individual bubble definitions (\emph{heterogenous fundamentals}; Section~\ref{sec_nonuni}). 

The node \emph{additional price properties} (see again Section~\ref{sec_def_price}) collects all papers which, in addition to the use of a definition of a fundamental value, relies on price properties for defining the bubble. For example, a price can be defined as deviation from a fundamental value which in addition has to have a boom bust dynamic. 

The next sub-category in the category \emph{with fundamentals} (see Section~\ref{sec_def_fund}) shows the different ways in which fundamental values are defined. These can rely on \emph{expected future cash flows} as in Equation \eqref{eq:definition_FV} or \eqref{eq:definition_FV_math} (see Section~\ref{par_cashflow}) or by the use of \emph{actual cash flows} (applied to past data; see Section~\ref{sec_Actual}). Furthermore, the definition can contain \emph{utility} other than cash flows as explained in Section~\ref{sec:utility_other_than_cash}, be defined by \emph{heuristics} or the \emph{book value} as described in Sections~\ref{Heu} and \ref{book}, rely on equating \emph{demand and supply} curves (see Section~\ref{MarketEq}) which is, e.g., used in some housing price models (see Section~\ref{sec_HOUSE}), or use an \emph{exogenously given} fundamental value which does not have to be computed via the use of expected cash flows (see Section~\ref{sec_exo}). Lastly, there are some papers that use \emph{other} definitions of bubbles than those mentioned up to now (see Section~\ref{other}). Since in our review, all those papers have some idea of fundamentals in mind, we add them to a subcategory of ``with fundamentals'' \citep[cf.][]{Biondo:13}.

On the process of categorizing the papers, we have to clarify some important points: Most papers cannot be assigned clearly to one category, that means we use the category which fits the best (as it is often the case when assigning things to categories). Some papers lie in more than one child-category, e.g., papers that use more than one definition. Another point to mention is that there are, e.g., papers that theoretically show that a bubble definition that relies on fundamentals is equivalent to one that uses volatility only. Other papers use this equivalent property. Then, the original theoretical paper is assigned to both with and without fundamental, the paper which uses for some case study this result is assigned to price information only since it does not matter ``for this paper'' that the used definition originally builds upon a fundamental-value involving definition. Sometimes, remaining works are categorized as ``others'' and sometimes papers do not lie in any child. 

\subsection{Results}\label{sec_results} 

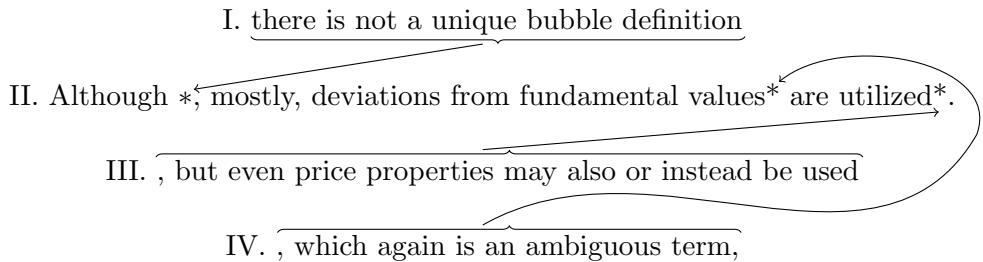
\begin{figure}\centering
\begin{tikzpicture}
\node(Punkt1) at(0,0){I. there is not a unique bubble definition};
\draw[decorate,decoration=brace] (3.45,-.2)--(-3.05,-.2);
\node at(0,-1){II. Although $*$, mostly, deviations from fundamental values* are utilized*.};
\draw[->](0,-.3)--(-3.8,-.9);
\node at(0,-2){III. , but even price properties may also or instead be used};
\draw[decorate,decoration=brace] (-4.3,-1.8)--(5,-1.8);
\draw[->](0,-1.7)--(6,-1.2);
\node at(0,-3){IV. , which again is an ambiguous term,};
\draw[decorate,decoration=brace] (-2.7,-2.8)--(3.4,-2.8);

\draw[->](0,-2.7)to[out=30,in=240](6.5,-1.5)to[out=70,in=40](3.9,-.8);

\end{tikzpicture}
\caption{The main result of the work at hand in one nested sentence. You may read only Item I.\ for the most important result, or Items I.\ and II.\ for the second result, or Items I., II., and III.\ for the third result, or, finally, all four items---if necessary, by combining the sentences as indicated by the stars and arrows.}\label{fig_result}
\end{figure}

Overall, we find that although there is not a unique bubble definition, mostly, deviations from fundamental values are utilized. In a minority of the papers, price properties may also or instead be used as definitional characteristics. Fundamental values in turn are not uniformly defined and operationalized across the literature. See Figure \ref{fig_result} for a summarizing presentation of the main result.
 
We furthermore present the results of the systematic review in more detail in the form of a tree schematic in Figure~\ref{tree2}. There, we provide the absolute number of papers in that category as well as the ratio of that node to all papers one node above. However, we note that the sum of the rations does not have to be equal to one in each case---rather smaller or larger values are possible due to the fact that some papers lie in no or in more than one child node. A full overview of the categorization results can be found in the supplementary material. Here, we are providing a summary of the categorization results. Recall that there are 122 (non-retracted) papers in the data set of the systematic review that fit the topic of financial bubbles. Regarding the areas of research in our sample, it is of no surprise that the majority of the papers are from finance~(27), followed by real estate~(23). While it might be quite natural that there are many papers from experimental economics~(16), macroecononics~(12), and commodities/agriculture/energy~(12), there are not so many papers in our review concerning bubbles in the fields of  heterogeneous agent models~(8), econophysics~(7), policy~(5), currency~(4), mathematical finance~(4), trade~(2), history of economics~(1), and geography~(1).

We continue with the question whether and what kind of definition is present.
There is one paper that does not fit the topic ``bubble,'' while it is still in the field of ``macro and general economics''. There are twelve papers without a  definition of what is meant by the term bubble but still fit the topic of bubbles, 25 papers where bubble is defined only verbally, twelve papers where we find both, definitions in words and definitions with a formula. Out of the papers with formulas which could be interpreted as bubble definition, 69 use a discrete time and 16 use continuous time formulations. That means, discrete time models seem to be widespread in academics in the understanding of bubbles. Note that all four mathematical finance papers are in continuous time. 
 
Coming to the main result, the vast majority~(92 out of 110, $84\%$) of the papers with definition use fundamentals (cf.\ Section~\ref{sec_def_fund})---where~73 use exclusively fundamental information and~19 use both price and fundamental information. Out of the papers using fundamentals, most (89) assume that fundamentals are unique while three  
 allow for heterogeneity. If fundamentals are allowed to be ambiguous, i.e.\ individual for each trader, it is for example assumed that a price is a bubble if it is above the supremum of all fundamentals. When fundamentals are used, quite often~(41) discounted expected future cash flows are utilized explicitly (see Section~\ref{sec_def_fund}). Several~(27)~papers use exogenously given fundamentals. Two papers utilize actual data (retrospective; see Section~\ref{sec_Actual}), six book values, and seven heuristics (see Sections~\ref{book} and~\ref{Heu}). 

In the cases when price path information is used to define a bubble, 31~papers use the price's rise, eleven 
crashes, and five utilize volatility as part of the bubble definition, cf.\ Section~\ref{sec_tests}. The last point might be especially interesting when thinking about market stability, bubbles, efficiency, and fluctuations, see Section~\ref{sec_Quotes}. We stress that the item ``heterogeneous fundamentals'' means that (a part of) the traders has/have heterogeneous (beliefs about the) fundamentals. Another reasonable thought one might have is that ``heterogeneous fundamentals'' means that the respective authors use different definitions of the term ``fundamental,'' but this is not how we apply it. Detailed results on the bubble definition categorization can be found in the supplementary material.

\subsection{Interpretation}\label{sec_inter}

\begin{figure}
\centering
\includegraphics[width=\textwidth]{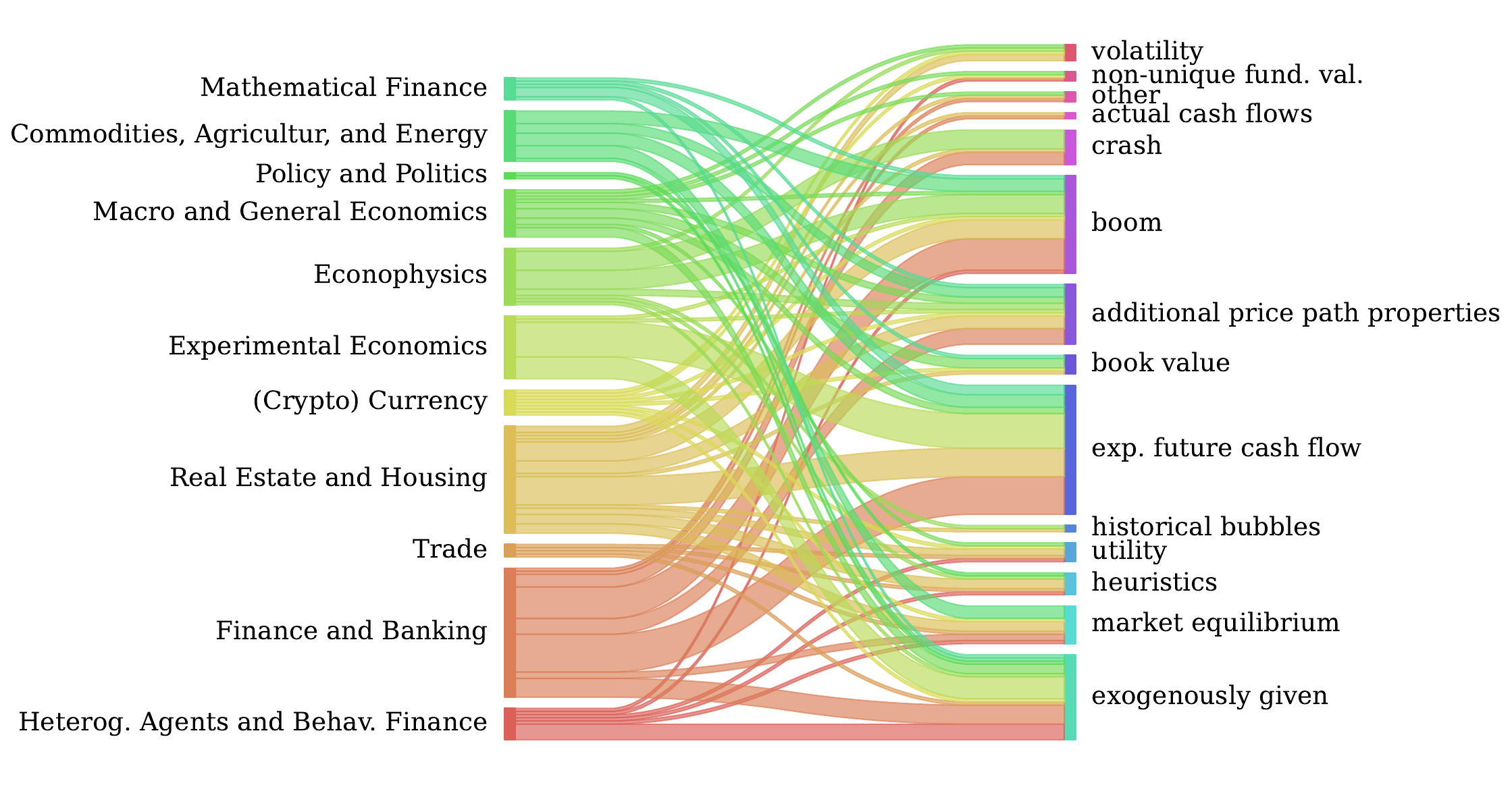}
\caption{Sankey plot, relating scientific disciplines/areas of research to the respective definition categories that the analysed studies fall into. We find that all research areas use a variety of definition types. (The plot was created using the Python package pySankey)}

\label{fig_sankey_research_areas}
\end{figure}
While one would expect that different research areas might use different definitions, we find a large discrepancy within research areas regarding the presented definition categories in our data set, see Figure \ref{fig_sankey_research_areas}. Especially in the larger definition categories such as ``boom", ``expected future cash flow" or ``exogenously given", we find almost all research areas represented. This means that there are no uniform definitions or operationalizations as categorized in the present study for each field. However, this does not mean that these different categories are incompatible. For example, using an exogenously given fundamental value might be a simplification of the concept of expected future cash flows. Several operationalizations that only use boom, bust, or price volatility properties in their definition are originally derived from an understanding of bubbles as deviations from fundamental values (see e.g. Section~\ref{sec_tests}). Thus, taking into account such derivations, the fraction of bubble definitions and operationalizations based on deviations from fundamentals is even larger than $84\%$, as found in the present data sample. 

On the other hand, due to the cutoff in the literature sample at a minimum of 15 citations, the data sample might have a bias towards more frequently used definition types, as papers with lower citations might use concepts that are less widely applied by other authors, due to a lower flow of information between the studies. This in turn might mean lower fractions of the prevalent definition categories in the bubble literature overall. 

We would like to point out that also within the defined categories, we find heterogeneity in definitions and operationalizations. For example, methods for detecting bubbles in real-time have been developed from different directions, leading to different operationalizations of bubbles. Bubble tests may identify exponentially rising dynamics, hyper-exponentially rising dynamics or certain properties of volatility (Section~\ref{sec_tests}). However, both exponentially rising and hyper-exponentially rising dynamics fall into the category of ``boom" although not being equivalent (Section~\ref{par_boom}).

While the idea of deviations from fundamentals gives rise to booms and bust cycles in case of stable fundamental values, boom and bust cycles do not necessarily imply deviations from fundamentals. If one focuses only on asset price dynamics as defining properties of bubbles, one cannot exclude the possibility that fundamental values behave in exactly the same fashion for the analysed period of time, leading to a phenomenon commonly known as a fad \citep[cf.][]{Camerer:89}. This also relates to the quote by Eugene Fama about the internet bubble (Section~\ref{sec_Quotes}), essentially meaning that expectations of future cash flows might have been rising as well during that time meaning growing fundamental values, i.e. not a bubble in terms of deviations from fundamentals. While in theory, it is possible to define a rational trader and derive a unique fundamental value with certain statistical properties (such that certain boom bust dynamics might be excluded), expectations in a colloquial sense of ``what people believe" are non-observable and differ across the population, and may well be behaving in a boom bust dynamic. Depending on whether we talk about theoretical rational expectations or empirical beliefs of people, the same category of definition, namely using expected future cash flows can lead to very different, non-equivalent outcomes. 

\subsection{Discussion}\label{sec_disc}
While individual theoretical definitions sometimes create an impression of clarity and unambiguity, various difficulties arise when moving from theory to real economic and financial systems. The following points are to be seen as inspiration for further discussions on defining bubbles.

Some widely spread definitions of bubbles give rise to problems such as circular statements in which the price and the intrinsic value of the asset depend on one another. Overall, the use of future cash flows in the definition of fundamentals can be difficult, if the asset does not yield to such cash flows. However, even if there are future cash flows, bubble definitions connected to them can be problematic as it is the case in real-estate prices and rents. If a buyer of a house assumes high rents, the calculated fundamental value of that object will be high, from which it follows that a high price is not a bubble. The other way around, when a buyer pays a high price, she or he wants to set high rents, to get her resp.\ his money back. That means, for deciding if high real-estate prices are a bubble, one needs to know whether rents are high resp.\ a bubble. For the determination whether rents are a bubble, one has to calculate opportunity costs---either by comparisons to other rents in that region or neighborhood, which will also be high if housing prices in that region are high due to renting and buying a house for living are partial substitutes,\footnote{We wish to thank Torsten Heinrich (TU Chemnitz) for this hint on substitutes.} or by comparison to credit rates for buying a house, which will in that case also be high. This, together, leads to a circular definition. 

A way out of the problem that there are assets without future cash flows and even without future opportunity costs could lie in the field of utility theory: one could define a bubble as a price that is so high that the expected future utility of keeping the money is (for all agents) higher than the expected future utility from buying that item. However, this leads to new problems, esp.\ when connecting it to the topic of finite vs.\ infinite horizons. When assuming that the utility of an item arises from having it (either via cash flows or other) up to a certain point in time and from reselling it at that point in time, two scenarios can happen. First, the reselling utility equals the utility from having it from that time up to eternity, and second, it does not. The second case could happen if a buyer paid at more than the infinite-horizon future utility justified. However, since she or he bought it, she or he assumed that the reselling utility was high enough to justify that price. This phenomenon is called a rational bubble---i.e., it is a bubble when only taking infinite horizons into account, but it is not one if allowing for reselling utility. If the so-called transversality condition is fulfilled, a price is also in the finite horizon definition not a bubble. However, if one paid to much once, it is rational for all future that traders pay to much (cf.\ birth of a bubble). The only case in that utility frame a bubble could happen is if one paid more than he or she expected to get back from all future utilities including reselling the so-called hot potato. 

There is the issue that fundamentals are subjective and latent, i.e.\ non observable. A way out is offered by retrospective definitions, which can be used to investigate whether past price dynamics were bubbles or not. In this case, one can use the real payoffs between the time of interest and today to calculate the fundamental value. However, these definitions not only have the problem that they are not directly applicable in practice, since one has to wait a long time \citep[about 30 years; cf.\ ``duration''; see][]{Siegel:03} to know whether a specific price movement was a bubble or not, but also a causality problem: if a price development is classified as a bubble in this point in time and thus policy measures are initiated, it may be that these measures lead to the fact that 30 years later the price development of that time would no longer be recognized as a bubble. However, this does not necessarily mean that policymakers should not have used the measures. This is a so-called self-defeating prophecy via the policy feedback loop: if a situation is identified as a bubble, and countermeasures are taken, it may (retrospectively) no longer be a bubble. Conversely, if a situation is not recognized as a starting bubble, it may become one.

The two definition possibilities of bubbles via fundamentals (via discounted expected future cash flows) and rising prices (looking at (mathematical) derivatives of price paths) are far less innocent than one might think. Imagine a price bubble in an asset, sector, or whole market. And imagine that the central bank decides to react to that. When interest rates are lowered, also the discounting factors in the definition of fundamentals are lower, leading to higher fundamentals that in turn justify high prices s.t.\ the bubble might have been gone. But when thinking about something like perfect or perfectly efficient markets, one might have financial markets in mind. Thus, traders are believed to anticipate the interest rate change such that not only fundamentals change, but also prices adjust. Some traders might trade in advance and some with delay, some too conservative and some too progressive, however, in the mean prices should adjust like fundamentals, thus, the interest rate change should have no consequence for this type of definition at all. And even in not-so-perfect markets, like real-estate, it would be hard to anticipate which one adjust faster and more: prices or fundamentals.---On the other hand, lowering interest rates pushes investors towards financial asset markets, raising demand, and in turn also raising prices and making the bubble even larger. When looking at price path slopes interest rates have to be increased to let the bubble burst. The question whether a central bank should have a look at price bubbles at all is complicated. Some central banks shall hold the consumer prices stable, where maybe rents are included, which depend on real estate prices. Thus, a housing bubble might be of interest to central banks, but with their absolute value and not with fundamentals as ratios. However, rent prices are often regulated, central bank interest rates are short-term rates and for discounting long term rates are used, and it is not clear whether risk-free rates or market rates shall be utilized, making things even more complicated.\footnote{The authors thank Jochen Hartwig (TU Chemnitz) for a very fruitful discussion on this topic.}

\section{Conclusion}\label{sec_con}
There is no consensus on what an asset price bubble is. Various definitions are used that refer only to market prices or only to fundamental values. We conducted a systematic review to assess which types of definitions are used how often. It turns out that definitions involving a fundamental value are widest spread. However, fundamental values bring the next problem, as they are not defined consistently either---and moreover, heuristics are sometimes taken as fundamental values. Although one can see that different branches or subfields of economics prefer certain definitions, e.g., via so-called fundamental values in the real estate field, it is not clear that fundamental values are dealt with in the same way in one field as in another. Interestingly, there is research linking definitions that only look at prices and those that also look at fundamentals. By defining how ``normal'' fundamental values behave, one can define bubbles or test for them when there is a price process that can never be justified by any fundamental values---such as hyper-exponential growth. Probably, the phenomenon ``bubble'' is high dimensional and, thus, hard to capture.

For future work it is important to analyze which fundamental value definitions are (in which circumstances) equivalent---and which price properties are (again: under which circumstances) equivalent to each other as well as to fundamental value types. The formulae for discounted values of cash flows with finite/infinite/indefinite (``stopping time'') and future/past/both data may be combined in one formula. Further, the presented manually created categorizations might serve as a learning data set for the application of machine learning algorithms in future projects. 

Since much discussion in economics could be avoided if people would talk about ``the same bubbles,'' our appeal is: Every researcher who works on the topic of bubbles should define very clearly and unambiguously what she understands by a bubble in her respective paper. We speculate that this is what \citet[][21:38--21:41/31:11]{Fama:13} intended to express by his somewhat exaggerated statement \begin{quote}`When people use the word ``bubble,'' they never tell you what they mean.'\end{quote}  given in his oral Nobel Prize Lecture \citep[cf.\ also][]{Engsted:16}. By stating clear and exact definitions it can be avoided that people talk past each other. 

\section*{DISCLOSURE STATEMENT}
The authors are not aware of any affiliations, memberships, funding, or financial holdings that
might be perceived as affecting the objectivity of this review. 

\section*{ACKNOWLEDGMENTS}
The authors wish to thank Lars Gr\"une and Bernhard Herz (both with University of Bayreuth), Torsten Heinrich and Jochen Hartwig (both with Chemnitz University of Technology), and Kerstin H\"otte (Oxford Martin School) for their helpful comments and suggestions. Further, the authors wish to thank Marina Eliza Spaliara (Adam Smith business school, University of Glasgow, Glasgow, Scotland, UK), who discussed this work at ICMAIF 2021 (see below).

This paper and its former versions were presented at AMASES Annual Conference, University of Reggio Calabria, Italy / online, 2021; International Conference on Macroeconomic Analysis and International Finance (ICMAIF), University of Crete, Rethymno, Greece / online, 2021; Maths in the Social Sciences Research Group, online, 2021; Summer School of the Tinbergen Institute Graduate School, Course `Behavioral Macro and Complexity,' 2021; European Association for Evolutionary Political Economy (EAEPE), Annual Conference, 2021.

This document was written in \LaTeX with help of Ti\textit{k}Z. Calculations were done in MS~Excel and Python.


\noindent
\begin{thebibliography}{00}

\bibitem[Abreu \& Brunnermeier(2003)]{Abreu:03}
Abreu D, Brunnermeier M. 2003.
Bubbles and Crashes.
\textit{Econometrica} 71(1):173--204 

\bibitem[Acharya \& Naqvi(2012)]{Acharya:12}
Acharya V, Naqvi H. 2012. 
The Seeds of a Crisis: A Theory of Bank Liquidity and Risk Taking over the Business Cycle. 
\textit{Journal of Financial Economics} 106(2):349--366 

\bibitem[Adland et~al.(2006)]{Adland:06}
Adland R, Haiying J, Strandenes S. 2006. 
Asset Bubbles in Shipping? An Analysis of Recent History in the Drybulk Market
\textit{Maritime Economics \& Logistics}
8:223--233

\bibitem[Allen et~al.(1993)]{Allen:93}
Allen F, Morris S, Postlewaite A. 1993.
Finite Bubbles with Short Sales Constraints and Asymmetric Information.
\textit{Journal of Economic Theory} 61(2):206--229

\bibitem[Allen \& Carletti(2013)]{Allen:13}
Allen F, Carletti E. 2013. Systemic risk from real estate and macro-prudential regulation
\textit{International Journal of Banking, Accounting and Finance} 5(1-2):28--48

\bibitem[Aliber \& Kindleberger(2015)]{Aliber:15}
Aliber RZ and Kindleberger CP. 2015. 
Manias, Panics, and Crashes: A History of Financial Crises. \textit{Palgrave Macmillan,} Basingstoke

\bibitem[Ambrose et~al.(2013)]{Ambrose:13}
Ambrose B W, Eichholtz P, Lindenthal T. 2013. 
House Prices and Fundamentals: 355 Years of Evidence. 
\textit{Journal of Money, Credit and Banking} 45(2-3):477--491

\bibitem[Barberis et~al.(2018)]{Barberis:18a}
Barberis N, Greenwood R, Jin L, Shleifer A. 2018. Extrapolation and bubbles. 
\textit{Journal of Financial Economics} 
129(2):203--227

\bibitem[Barberis(2018)]{Barberis:18b}
Barberis N. 2018. Psychology-based Models of Asset Prices and Trading Volume. 
\textit{In: Handbook of Behavioral Economics: Applications and Foundations 1}(1):79--175. North-Holland.

\bibitem[Barlevy(2007)]{Barlevy:07}
Barlevy G. 2007.
Economic Theory and Asset Bubbles.
\textit{Economic perspectives. Federal Reserve Bank of Chicago} Third Quater 2007:44--59

\bibitem[Ba\c{s}o\u{g}lu Kabran and \"Unl\"u(2021)]{Basoglue:21}
Ba\c{s}o\u{g}lu Kabran F and \"Unl\"u KD. 2021. 
A Two-Step Machine Learning Approach to Predict S\&P 500 Bubbles.
\textit{Journal of Applied Statistics} 48(13-15):2776--2794

\bibitem[Baur \& Glover(2014)]{Baur:14}
Baur D G, Glover K J. 2014. Heterogeneous Expectations in the Gold Market: Specification and Estimation. 
\textit{Journal of Economic Dynamics and Control} 40:116--133

\bibitem[Beja \& Goldman(1980)]{Beja:80}
Beja A and Goldman MB. 1980. On the Dynamic Behavior of Prices in Disequilibrium.
\textit{The Journal of Finance} 35(2):235--248

\bibitem[Biagini et~al.(2014)]{Biagini:14}
Biagini F, F\"ollmer H, Nedelcu S. 2014.
Shifting Martingale Measures and the Birth of a Bubble as a Submartingale. 
\textit{Finance and Stochastics} 18(2):297--326

\bibitem[Biagini \& Nedelcu(2015)]{Biagini:15}
Biagini F, Nedelcu S. 2015.
The Formation of Financial Bubbles in Defaultable Markets.
\textit{SIAM Journal on Financial Mathematics} 6(1):530--558

\bibitem[Bia\l kowski et~al.(2015)]{Bialkowski:15}
Bia\l kowski J, Bohl MT, Stephan PM, Wisniewski TP. 2015.
The Gold Price in Times of Crisis.
\textit{International Review of Financial Analysis}
41:329--339

\bibitem[Biondo et~al.(2013)]{Biondo:13}
Biondo AE, Pluchino A, Rapisarda A, Helbing D. 2013. 
Reducing Financial Avalanches by Random Investments. 
\textit{Physical Review E} 88(6):062814

\bibitem[Blanchard \& Watson(1982)]{Blanchard:82}
Blanchard OJ, Watson MW. 1982.
Bubbles, Rational Expectations and Financial Markets. 
\textit{NBER Working Paper} w0945

\bibitem[Bourassa et~al.(2001)]{Bourassa:01}
Bourassa S C, Hendershott, P H, Murphy J. 2001. 
Further Evidence on the Existence of Housing Market Bubbles. 
\textit{Journal of Property Research} 18(1):1--19

\bibitem[Br\'ee \& Joseph(2013)]{Bree:13}
Br\'ee D S, Joseph N L. 2013. 
Testing for Financial Crashes Using the Log Periodic Power Law Model. 
\textit{International Review of Financial Analysis} 30:287--297

\bibitem[Brunnermeier \& Oehmke(2013)]{Brunnermeier:13}
Brunnermeier M K, Oehmke M. 2013. Bubbles, financial crises, and systemic risk. 
\textit{Handbook of the Economics of Finance} 2: 1221--1288

\bibitem[Brzezicka(2020)]{Brzezicka:20}
Brzezicka J. 2020.
Towards a Typology of Housing Price Bubbles: A Literature Review. 
\textit{Housing, Theory and Society} 1--23

\bibitem[Caballero \& Krishnamurthy(2006)]{Caballero:06}
Caballero R, Krishnamurthy A. 2006.
Bubbles and Capital Flow Volatility: Causes and Risk Management.
\textit{Journal of Monetary Economy} 53(1):35--53

\bibitem[Cajueiro et~al.(2009)]{Cajueiro:09}
Cajueiro D O, Tabak B M, Werneck F K. 2009. 
Can We Predict Crashes? The Case of the Brazilian Stock Market.
\textit{Physica A: Statistical Mechanics and its Applications} 388(8):1603--1609

\bibitem[Camerer(1989)]{Camerer:89}
Camerer C. 1989.
Bubbles and Fads in Asset Prices.
\textit{Journal of Economic Surveys} 3(1):3--41

\bibitem[Campbell et~al.(2009)]{Campbell:09}
Campbell S D, Davis M A, Gallin J, Martin R F. 2009. 
What Moves Housing Markets: A Variance Decomposition of the Rent-Price Ratio. 
\textit{Journal of Urban Economics} 66(2):90--102

\bibitem[Carter et~al.(2011)]{Carter:11}
Carter C A, Rausser G C, Smith A. 2011.
Commodity Booms and Busts.
\textit{Annual Review of Resource Economics} 3(1):87--118

\bibitem[Cheung et al.(2012)]{Cheung:12}
Cheung S L, Palan, S. 2012. Two heads are less bubbly than one: team decision-making in an experimental asset market. \textit{Experimental Economics} 15(3):373--397

\bibitem[Chiarella et~al.(2009)]{Chiarella:09}
Chiarella C, Dieci R, He X Z. 2009. 
Heterogeneity, Market Mechanisms, and Asset Price Dynamics. 
\textit{Handbook of financial markets: Dynamics and evolution, North-Holland} 277--344

\bibitem[Clark et~al.(2010)]{Clark:10}
Clark G L, Durán-Fernández R, Strauss K. 2010. 
‘Being in the market’: the UK house-price bubble and the intended structure of individual pension investment portfolios. 
\textit{Journal of Economic Geography} 10(3):331--359

\bibitem[Clement(2007)]{Clement:07}
Clement D. 2007. 
Interview with Eugene Fama. 
\textit{The Region}; 2~November~2007 resp.\ 1~December~2007 \url{https://www.minneapolisfed.org/article/2007/interview-with-eugene-fama} Checked: 20~September~2022

\bibitem[Cogley(1999)]{Cogley:99}
Cogley T. 1999.
Should the Fed take Deliberate Steps to Deflate Asset Price Bubbles?
\textit{Economic Review. Federal Reserve Bank of San Francisco} 1:42--52

\bibitem[Conlon(2004)]{Conlon:04}
Conlon J. 2004.
Simple Finite Horizon Bubbles Robust to Higher Order Knowledge.
\textit{Econometrica} 72(3):927--936

\bibitem[Cox \& Hobson(2005)]{Cox:05}
Cox A, Hobson D G. 2005. 
Local Martingales, Bubbles and Option Prices. 
\textit{Finance and Stochastics} 9(4):477--492

\bibitem[Criens(2020a)]{Criens:20a}
Criens D. 2020.
No Arbitrage in Continuous Financial Markets.
\textit{Mathematics and Financial Economics} 14:461--506

\bibitem[Criens(2020b)]{Criens:20b}
Criens DBM. 2020.
Essays on Stochastic Processes and their Applications.
\textit{Doctoral thesis, Technische Universit\"at M\"unchen }

\bibitem[Day \& Huang(1990)]{Day:90}
Day R H, Huang W. 1990. Bulls, Bears and Market Sheep. \textit{Journal of Economic Behavior \& Organization} 14(3):299--329

\bibitem[Delbaen \& Schachermayer(1994)]{Delbaen:94}
Delbaen F, Schachermayer W. 1994.
A General Version of the Fundamental Theorem of Asset Pricing. 
\textit{Mathematical Annals} 300:463--520

\bibitem[Delbaen \& Schachermayer(1998)]{Delbaen:98}
Delbaen F, Schachermayer W. 1998.
The Fundamental Theorem of Asset Pricing for Unbounded Stochastic Processes. 
\textit{Mathematical Annals} 312:215--250

\bibitem[Delbaen \& Schachermayer(2004)]{Delbaen:04}
Delbaen F, Schachermayer W. 2004.
What is \dots\ a Free Lunch?
\textit{Notices of the AMS} 51(5):526--528

\bibitem[De Long et~al.(1990)]{DeLong:90}
De Long JB, Shleifer A, Summers L, Waldmann R. 1990.
Noise Trader Risk in Financial Markets.
\textit{Journal of Political Economy} 98(4):703--738

\bibitem[Diamond(1965)]{Diamond:65}
Diamond P. 1965.
National Debt in a Neoclassical Growth Model. Part 1. \textit{American Economic Review} 55(5):1126--1150

\bibitem[Diba \& Grossman(1988)]{Diba:88}
Diba BT, Grossman HI. 1988. 
The Theory of Rational Bubbles in Stock Prices. 
\textit{The Economic Journal} 98(392):746--754

\bibitem[Doblas-Madrid(2012)]{Doblas:12}
Doblas-Madrid A. 2012.
A Robust Model of Bubbles with Multidimensional Uncertainty.
\textit{Econometrica} 80(5):1845--1883

\bibitem[Downarowicz(2010)]{Downarowicz:10}
Downarowicz A. 2010. The First Fundamental Theorem of Asset Pricing. 
\textit{Revista de Econom\'{i}a Financiera} 21:23--35

\bibitem[Dreger \& Zhang(2013)]{Dreger:13}
Dreger C, Zhang Y. 2013. 
Is there a bubble in the Chinese housing market? 
\textit{Urban Policy and Research} 31(1):27--39

\bibitem[Elsner et al.(2014)]{Elsner:14}
Elsner W, Heinrich T, Schwardt H. 2014. 
The microeconomics of complex economies: Evolutionary, institutional, neoclassical, and complexity perspectives. \textit{Academic Press.}

\bibitem[Engsted(2016)]{Engsted:16}
Engsted T. 2016.
Fama on Bubbles.
\textit{Journal of Economic Surveys} 30(2):370--376

\bibitem[Evans(1986)]{Evans:86}
Evans G W. 1986. 
A test for speculative bubbles in the sterling-dollar exchange rate: 1981-84. 
\textit{The American Economic Review} 621--636

\bibitem[Fama(1970)]{Fama:70}
Fama E F. 1970. Efficient Capital Markets: A Review of Theory and Empirical Work. 
\textit{Journal of Finance} 25(2):383--417

\bibitem[Fama(2013)]{Fama:13}
Fama EF. 2013.
Price Lecture.
\textit{The Nobel Prize} \url{https://www.nobelprize.org/prizes/economic-sciences/2013/fama/lecture/}

\bibitem[Fama(2014)]{Fama:14}
Fama E F. 2014. Two pillars of asset pricing. 
\textit{American Economic Review} 104(6):1467--1485

\bibitem[Farhi \& Tirole(2012)]{Farhi:12}
Farhi E, Tirole J. 2012.
Bubbly Liquidity.
\textit{Review of Economic Studies} 79(2):678--706

\bibitem[Franke \& Westerhoff(2016)]{Franke:16}
Franke R, Westerhoff F. 2016. Why a Simple Herding Model may Generate the Stylized Facts of Daily Returns: Explanation and Estimation. 
\textit{Journal of Economic Interaction and Coordination} 11(1):1--34

\bibitem[Friede et~al.(2015)]{Friede:15}
Friede G, Busch T, Bassen A. 2015. 
ESG and Financial Performance: Aggregated Evidence from more than 2000 Empirical Studies. 
\textit{Journal of Sustainable Finance \& Investment} 5(4):210--233

\bibitem[Friedman \& Abraham(2009)]{Friedman:09}
Friedman D, Abraham R. 2009.
Bubbles and Crashes: Gradient Dynamics in Financial Markets.
\textit{Journal of Economic Dynamics and Control} 33(4):922--937

\bibitem[Fry \& Cheah(2016)]{Fry:16}Fry J, Cheah E-T. 2016. Negative Bubbles and Shocks in Cryptocurrency Markets. \textit{International Review of Financial Analysis} 47:343--352

\bibitem[Gallant(2021)]{Investopedia1}
Gallant C. 2021. How Are Book Value and Intrinsic Value Different? \textit{Investopedia,} \url{https://www.investopedia.com/ask/answers/05/bookvintrinsic.asp}, checked: 2024-02-20

\bibitem[Garber(2001)]{Garber:01}
Garber PM. 2001. 
Famous First Bubbles: The Fundamentals of Early Manias. \textit{The MIT Press,} Cambridge, Massachusetts, London, England

\bibitem[Girdzijauskas \& \v{S}treimikien\.{e}(2009)]{Girdzijauskas:09}
Girdzijauskas S, \v{S}treimikien\.{e} D. 2009. 
Application of Logistic Models for Stock Market Bubbles Analysis. 
\textit{Journal of Business Economics and Management} 10(1):45--51

\bibitem[Glaeser \& Nathanson(2015)]{Glaeser:15}
Glaeser E L, Nathanson C G. 2015. 
Housing Bubbles. 
In: \textit{Handbook of regional and urban economics, Elsevier} 5:701--751

\bibitem[Greenwood et~al.(2019)]{Greenwood:19}Greenwood R, Shleifer A, You Y. 2019. Bubbles for Fama. \textit{Journal of Financial Economics} 131:20--43

\bibitem[Guasoni et~al.(2010)]{Guasoni:10}
Guasoni P, R\'{a}sonyi M, Schachermayer W. 2010. 
The Fundamental Theorem of Asset Pricing for Continuous Processes under Small Transaction Costs. 
\textit{Annals of Finance} 6(2):157--191

\bibitem[Gusenbauer \& Haddaway(2020)]{Gusenbauer:20}
Gusenbauer M, Haddaway N R. 2020. 
Which academic search systems are suitable for systematic reviews or meta-analyses? Evaluating retrieval qualities of Google Scholar, PubMed, and 26 other resources. \textit{Research synthesis methods} 11(2):181--217

\bibitem[G\"urkaynak(2008)]{Gurkaynak:08}
G\"urkaynak R S. 2008. Econometric tests of asset price bubbles: taking stock. \textit{Journal of Economic surveys} 22(1):166--186

\bibitem[Harras \& Sornette(2011)]{Harras:11}
Harras G, Sornette D. 2011. How to grow a bubble: A model of myopic adapting agents. \textit{Journal of Economic Behavior \& Organization} 80(1):137--152

\bibitem[Haruvy et~al.(2007)]{Haruvy:07}
Haruvy E, Lahav Y, Noussair C. 2007. Traders' expectations in asset markets: experimental evidence. \textit{American Economic Review} 97(5):1901--1920

\bibitem[Hayashi(1982)]{Hayashi:82}
Hayashi F. 1982. 
Tobin's Marginal q and Average q: A Neoclassical Interpretation. \textit{Econometrica: Journal of the Econometric Society} 50(1):213--224

\bibitem[Hayes(2023)]{Investopedia2}
Hayes A. 2023. Book Value: Definition, Meaning, Formula, and Examples. \textit{Investopedia,} \url{https://www.investopedia.com/terms/b/bookvalue.asp}, checked: 2024-02-20

\bibitem[Hirota \& Sunder(2016)]{Hirota:16}
Hirota S, Sunder S. 2016. Price bubbles sans dividend anchors: Evidence from laboratory stock markets. 
In: \textit{Behavioral Interactions, Markets, and Economic Dynamics} Springer, Tokyo. 357--395 

\bibitem[Hommes(2006)]{Hommes:06}
Hommes C. 2006. Heterogeneous Agent Models in Economics and Finance. 
\textit{Handbook of Computational Economics} 2:1109--1186

\bibitem[Hommes(2021)]{Hommes:21}
Hommes C. 2021. Behavioral and experimental macroeconomics and policy analysis: A complex systems approach. 
\textit{Journal of Economic Literature} 59(1):149--219

\bibitem[Horst(2005)]{Horst:05}
Horst U. 2005. Financial price fluctuations in a stock market model with many interacting agents. 
\textit{Economic Theory} 25(4):917--932

\bibitem[Hott(2009)]{Hott:09}
Hott C. 2009. Herding behavior in asset markets. 
\textit{Journal of Financial Stability} 5(1):35--56

\bibitem[Jarrow et~al.(2010)]{Jarrow:10}
Jarrow RA, Protter P, Shimbo K. 2010. 
Asset Price Bubbles in Incomplete Markets.
\textit{Mathematical Finance: An International Journal of Mathematics, Statistics and Financial Economics} 20(2):145--185

\bibitem[Jarrow \& Larsson(2012)]{Jarrow:12a}
Jarrow RA, Larsson M. 2012.
The Meaning of Market Efficiency.
\textit{Mathematical Finance} 22(1):1--30

\bibitem[Jarrow et al.(2012)]{Jarrow:12b}
Jarrow R A, Protter P, Roch A F. 2012. 
A liquidity-based model for asset price bubbles. 
\textit{Quantitative Finance} 12(9):1339--1349

\bibitem[Jarrow(2015)]{Jarrow:15}
Jarrow RA. 2015. 
Asset Price Bubbles. 
\textit{Annual Review of Financial Economics} 7:201--218

\bibitem[Jarrow(2016)]{Jarrow:16}
Jarrow RA. 2016. 
Testing for Asset Price Bubbles: Three New Approaches.  
\textit{Quantitative Finance Letters} 4(1):4--9

\bibitem[Jiang et~al.(2010)]{Jiang:10}
Jiang Z-Q, Zhou W-X, Sornette D, Woodard R, Bastiaensen K, Cauwels P. 2010. 
Bubble diagnosis and prediction of the 2005-2007 and 2008-2009 Chinese stock market bubbles. 
\textit{Journal of economic behavior \& organization}, 74(3):149--162


\bibitem[Johansen(2003)]{Johansen:03}
Johansen A. Characterization of Large Price Variations in Financial Markets. 
\textit{Physica A} 324:157--166

\bibitem[Johansen et~al.(2000)]{Johansen:00}
Johansen A, Ledoit O, Sornette D. 2000. 
Crashes as critical points. 
\textit{International Journal of Theoretical and Applied Finance} 3(02):219--255

\bibitem[Juvenal \& Petrella(2015)]{Juvenal:15}
Juvenal L, Petrella I. 2015. Speculation in the oil market. \textit{Journal of applied econometrics}
30(4):621--649

\bibitem[Kaizoji(2000)]{Kaizoji:00}Kaizoji T. 2000. Speculative Bubbles and Crashes in Stock Markets: An Interacting-Agent Model of Speculative Activity. \textit{Physica A} 287:493--506

\bibitem[Kindleberger(1978)]{Kindleberger:78}
Kindleberger CP. 1978.
Manias, Panics, and Crashes
\textit{New York: Basic Books}

\bibitem[Kindleberger(1996)]{Kindleberger:96}
Kindleberger CP. 1996.
Manias, Panics, and Crashes: A History of Financial Crises
\textit{Wiley}

\bibitem[Kindleberger \& Aliber(2005)]{Kindleberger:05}
Kindleberger CP, Aliber RZ. 2005.
Manias, Panics, and Crashes: A History of Financial Crises
\textit{Wiley}

\bibitem[Aliber et~al.(2023)]{Kindleberger:23}
Aliber RZ, Kindleberger CP, RN McCauley. 2023.
Manias, Panics, and Crashes: A History of Financial Crises
\textit{Wiley}

\bibitem[Komarek \& Kubicov\'{a}(2011)]{Kubicova:11}
Komarek L, Kubicov\'{a} I. 2011. 
The Classification and Identification of Asset Price Bubbles. 
\textit{Czech Journal of Economics and Finance (Finance a Uver)} 61(1):34--48

\bibitem[Kyriazis et~al.(2020)]{Kyriazis:20}
Kyriazis N, Papadamou S, Corbet S. 2020. 
A Systematic Review of the Bubble Dynamics of Cryptocurrency Prices. 
\textit{Research in International Business and Finance} 101254

\bibitem[Lei et~al.(2001)]{Lei:01}
Lei V, Noussair C N, Plott C R. 2001. Nonspeculative bubbles in experimental asset markets: Lack of common knowledge of rationality vs. actual irrationality. \textit{Econometrica} 69(4):831--859

\bibitem[Lee et~al.(1999)]{Lee:99}
Lee CM, Myers J, Swaminathan B. 1999. 
What is the Intrinsic Value of the Dow?
\textit{The Journal of Finance} 54(5):1693--1741

\bibitem[Loewenstein \& Willard(2000)]{Loewenstein:00}
Loewenstein M, Willard GA. 2000.
Rational Equilibrium Asset-Pricing Bubbles in Continuous Trading Models.
\textit{Journal of Economic Theory} 91(1):17--58

\bibitem[Martin \& Ventura(2018)]{Martin:18}
Martin A, Ventura J. 2018. The Macroeconomics of Rational Bubbles: A User's Guide. 
\textit{Annual Review of Economics} 10(1):505--539

\bibitem[Malkiel(1989)]{Malkiel:89}
Malkiel BG. 1989. Efficient Market Hypothesis. \textit{In: Eatwell J, Milgate M, Newman P (eds): Finance. The New Palgrave. Palgrave Macmillan, London}

\bibitem[Malkiel(2005)]{Malkiel:05}
Malkiel BG. Reflections on the Efficient Market Hypothesis: 30 Years Later. \textit{Financial
Review}, 40(1):1--9

\bibitem[Mas-Colell et al.(1995)]{Mas-colell:95}
Mas-Colell A, Whinston M D, Green J R. 1995. Microeconomic theory (Vol. 1). 
\textit{New York: Oxford university press.}

\bibitem[Miao et al.(2015)]{Miao:15}
Miao J, Wang P, Xu Z. 2015. 
A Bayesian dynamic stochastic general equilibrium model of stock market bubbles and business cycles. \textit{Quantitative Economics,} 6:599--635

\bibitem[Milgrom \& Stokey(1982)]{Milgrom:82}
Milgrom P, Stokey N. 1982.
Information, Trade, and Common Knowledge.
\textit{Journal of Economic Theory} 26(1):991--1009

\bibitem[Noussair et~al.(2001)]{Noussair:01}
Noussair C N, Robin S, Ruffieux B. 2001. Price bubbles in laboratory asset markets with constant fundamental values. \textit{Experimental Economics} 4:87--105 

\bibitem[Ozgur et~al.(2021)]{Ozgur:21}
Ozgur O, Yilanci V, Ozbugday FC. 2021. 
Detecting Speculative Bubbles in Metal Prices: Evidence from GSADF Test and Machine Learning Approaches. \textit{Resources Policy} 74:102306

\bibitem[Page et~al.(2021)]{Page:21}
Page M J, McKenzie J E, Bossuyt P M, Boutron I, Hoffmann T C, Mulrow C D, ... and Moher D. 2021. The PRISMA 2020 statement: an updated guideline for reporting systematic reviews. \textit{bmj}, 372

\bibitem[Palan(2013)]{Palan:13}
Palan S. 2013. A review of bubbles and crashes in experimental asset markets. \textit{Journal of Economic surveys} 27(3): 570--588

\bibitem[Phillips et~al.(2011)]{PhillipsWuYu:11}
Phillips P, Wu Y, Yu J. 2011. 
Explosive behavior in the 1990s Nasdaq, when did exuberance escalate asset values? 
\textit{International Economic Review} 52:201--206

\bibitem[Phillips \& Yu(2011)]{PhillipsYu:11}
Phillips PC, Yu J. 2011. 
Dating the timeline of financial bubbles during the subprime crisis. 
\textit{Quantitative Economics} 2(3):455--491

\bibitem[Phillips et~al.(2012)]{Phillips:12}
Phillips P C, Shi S, Yu J. 2012. Testing for multiple bubbles.

\bibitem[Phillips et~al.(2015)]{Phillips:15}
Phillips P C, Shi S, Yu J. 2015. 
Testing for multiple bubbles: Historical episodes of exuberance and collapse in the S$\&$P 500. 
\textit{International economic review} 56(4):1043--1078

\bibitem[Platen \& Tappe(2020)]{Platen:20}
Platen E, Tappe S. 2020. 
The Fundamental Theorem of Asset Pricing for Self-Financing Portfolios.
\textit{arXiv preprint}

\bibitem[Platen \& Tappe(2021)]{Platen:21}
Platen E, Tappe S. 2021. 
No Arbitrage and Multiplicative Special Semimartingales
\textit{arXiv preprint}

\bibitem[Protter(2016)]{Protter:16}
Protter P. 2016. 
Bubbles and Crashes -- Mathematical Models of Bubbles.
\textit{Quantitative Finance Letters} 4(1):10--13

\bibitem[Rocheteau \& Wright(2011)]{Rocheteau:11}
Rocheteau G, Wright R. 2011.
Liquidity and Asset Market Dynamics.
\textit{Mimeo, University of Wisconsin}

\bibitem[Samuelson(1958)]{Samuelson:58}
Samuelson P. 1958.
An Exact Consumption-Loan Model of Interest with or without the Social Contrivance of Money. \textit{Journal of Political Economy} 66:467--482

\bibitem[Schatz \& Sornette(2020)]{Schatz:20}
Schatz M, Sornette D. 2020. 
Inefficient bubbles and efficient drawdowns in financial markets. 
\textit{International Journal of Theoretical and Applied Finance} 23(07):2050047

\bibitem[Shiller(2014)]{Shiller:14}
Shiller R J. 2014. Speculative asset prices. 
\textit{American Economic Review} 104(6):1486--1517

\bibitem[Siegel(2003)]{Siegel:03}
Siegel JJ. 2003.
What Is an Asset Price Bubble? An Operational Definition.
\textit{European Financial Management} 9:11--24

\bibitem[Sin(1996)]{Sin:96}
Sin CA. 1996.
Strictly Local Martingales and Hedge Ratios on Stochastic Volatility Models.
\textit{PhD thesis, Cornell University}

\bibitem[Sornette et~al.(1996)]{Sornette:96}
Sornette D, Johansen A, Bouchaud J P. 1996. 
Stock market crashes, precursors and replicas. 
\textit{Journal de Physique I} 6(1):167--175

\bibitem[Sornette(2003)]{Sornette:03}
Sornette, D. 2003. Critical market crashes. 
\textit{Physics Reports} 378(1):1--98

\bibitem[Sornette \&Cauwels(2014)]{Sornette:14}Sornette D, Cauwels P. 2014. 
1980--2008: The Illusion of the Perpetual Money Machine and What it Bodes for the Future. 
\textit{Risks} 2:103--131

\bibitem[Sornette et~al.(2018)]{Sornette:18}
Sornette D, Cauwels P, Smilyanov G. 2018. 
Can We Use Volatility to Diagnose Financial Bubbles? Lessons from 40 Historical Bubbles. 
\textit{Quantitative Finance and Economics} 2(1):486--594

\bibitem[St\"ockl et~al.(2010)]{Stoeckl:10}
St\"ockl T, Huber J, Kirchler M. 2010.
Bubble Measures in Experimental Asset Markets.
\textit{Experimental Economics} 13(3):284--298

\bibitem[Summers(1986)]{Summers:86}
Summers L H. 1986. 
Does the stock market rationally reflect fundamental values? 
\textit{The Journal of Finance} 41(3), 591--601

\bibitem[Thurner et~al.(2012)]{Thurner:12}
Thurner S, Farmer JD, Geanakoplos J. 2012. 
Leverage Causes Fat Tails and Clustered Volatility. 
\textit{Quantitative Finance} 12(5):695--707

\bibitem[Tirole(1982)]{Tirole:82}
Tirole J. 1982.
On the Possibility of Speculation under Rational Expectations.
\textit{Econometrica} 50(5):1163--1182

\bibitem[Tirole(1985)]{Tirole:85}
Tirole J. 1985.
Asset Bubbles and Overlapping Generations. \textit{Econometrica} 53(6):1499--1528

\bibitem[Tobin(1969)]{Tobin:69}
Tobin J. 1969. 
A General Equilibrium Approach to Monetary Theory. \textit{Journal of Money, Credit and Banking,} Ohio State University Press, 1(1):15--29

\bibitem[Townsend(1980)]{Townsend:80}
Townsend B. 1980.
Models of Money with Spatially Separated Agents
\textit{Models of Monetary Economies. Federal Reserve Bank of Minneapolis} 

\bibitem[van Zanten(2008)]{Zanten:08}
van Zanten H. 2008.
Lectures on the FTAP---Notes Complementing Delbaen and
Schachermayer's Book ``The Mathematics of Arbitrage''

\bibitem[Vissing-Jorgensen(2003)]{Vissing:03}
Vissing-Jorgensen A. 2003. 
Perspectives on behavioral finance: Does "irrationality" disappear with wealth? Evidence from expectations and actions. 
\textit{NBER macroeconomics annual}, 18:139--194

\bibitem[Vogel(2018)]{Vogel:18}
Vogel HL. 2018. 
Financial Market Bubbles and Crashes: Features, Causes, and Effects. \textit{Springer}

\bibitem[Werner(1997)]{Werner:97}
Werner J. 1997. 
Arbitrage, Bubbles, and Valuation.
\textit{International Economic Review} 453--464 

\bibitem[Wilson(1997)]{Wilson:97}
Wilson R. (1997). 
Islamic Finance and Ethical Investment. \textit{International Journal of Social Economics}

\bibitem[Yan et~al.(2010)]{Yan:10}
Yan W, Woodard R, Sornette D. 2010. 
Diagnosis and Prediction of Tipping Points in Financial Markets: Crashes and Rebounds. 
\textit{Physics Procedia} 3(5):1641--1657

\bibitem[Yiu et~al.(2013)]{Yiu:13}
Yiu M S, Yu J, Jin L. 2013. 
Detecting Bubbles in Hong Kong Residential Property Market. 
\textit{Journal of Asian Economics} 28:115--124

\bibitem[Youssefmir et~al.(1998)]{Youssefmir:98}
Youssefmir M, Huberman B A, Hogg T. 1998. 
Bubbles and Market Crashes. 
\textit{Computational Economics} 12(2):97--114

\bibitem[Zeeman(1974)]{Zeeman:74}
Zeeman E C. 1974. 
On the Unstable Behaviour of Stock Exchanges. 
\textit{Journal of mathematical economics} 1(1):39--49

\bibitem[Zeng(2009)]{Zeng:09}
Zeng Y. 2009.
Fundamental Theorem of Asset Pricing in a Nutshell: With a View toward Num\'{e}raire Change.

\bibitem[Zhang et~al.(2016)]{Zhang:16}
Zhang Q, Sornette D, Balcilar M, Gupta R, Ozdemir Z A, Yetkiner H. 2016. 
LPPLS Bubble Indicators over two Centuries of the S\& P 500 Index. 
\textit{Physica A: statistical Mechanics and its Applications} 458:126--139
 
\bibitem[Zhang \& Wang(2015)]{Zhang:15}
Zhang Y J, Wang J. 2015. 
Exploring the WTI Crude Oil Price Bubble Process using the Markov Regime Switching Model. 
\textit{Physica A: Statistical mechanics and its applications},
421:377--387

\bibitem[Zhang \& Yao(2016)]{ZhangYao:16}
Zhang Y J, Yao T. 2016. Interpreting the movement of oil prices: driven by fundamentals or bubbles? 
\textit{Economic Modelling} 55:226--240 

\end{thebibliography}
\end{document}